\documentclass[fleqn,usenatbib,useAMS]{mnras}

\usepackage[dvipsnames]{xcolor}
\usepackage{tikz,hyperref}

\definecolor{lime}{HTML}{A6CE39}
\DeclareRobustCommand{\orcidicon}{
	\begin{tikzpicture}
	\draw[lime, fill=lime] (0,0) 
	circle [radius=0.13] 
	node[white] {{\fontfamily{qag}\selectfont \tiny ID}};
	\draw[white, fill=white] (-0.0625,0.095) 
	circle [radius=0.007];
	\end{tikzpicture}
	\hspace{-2mm}
}

\foreach \x in {A, ..., Z}{\expandafter\xdef\csname orcid\x\endcsname{\noexpand\href{https://orcid.org/\csname orcidauthor\x\endcsname}
			{\noexpand\orcidicon}}
}

\usepackage{newtxtext,newtxmath}

\usepackage{booktabs}

\usepackage[T1]{fontenc}
\usepackage{ae,aecompl}

\usepackage{graphicx}	
\usepackage{amsmath}	
\usepackage{booktabs}
\usepackage{times}
\input{epsf}

\title[The X-ray properties of Symbiotic Systems]{The Influence of the Accretion Disc Structure on X-ray Spectral States in Symbiotic Binaries}

\author[Jes\'{u}s~A.~Toal\'{a} \& Diego~A.~Vasquez-Torres]{Jes\'{u}s~A.~Toal\'{a}\thanks{E-mail:\,j.toala@irya.unam.mx}$^{1,2}\orcidA$ and Diego~A.~Vasquez-Torres$^{1}\orcidB$\\ 
$^{1}$Instituto de Radioastronom\'{i}a y Astrof\'{i}sica, Universidad Nacional Aut\'{o}noma de M\'{e}xico, Morelia 58089, Michoac\'{a}n, Mexico\\
$^{2}$Facultad de Ciencias de la Tierra y el Espacio, Universidad Aut\'{o}noma de Sinaloa, Josefa Ortiz de Dom\'{i}nguez S/N, Culiac\'{a}n 80040, Sin., Mexico
}

\date{\today}

\pubyear{2025}

\begin{document}
\label{firstpage}
\pagerange{\pageref{firstpage}--\pageref{lastpage}}
\maketitle

\begin{abstract}
Symbiotic stars are binary systems where a white dwarf (WD) accretes material from the wind of an evolved, late-type companion. X-ray-emitting symbiotic systems are classified into $\alpha$, $\beta$, $\delta$, and $\beta/\delta$ types, attributed to distinct physical mechanisms such as thermonuclear burning, wind interactions, and accretion-driven boundary layers. We present synthetic X-ray spectra derived from hydrodynamics simulations using the {\sc phantom} code, coupled with radiative-transfer calculations from {\sc skirt}. We reproduce all X-ray spectral types by exploring different density structure of the accretion disc, the viewing angle, the plasma temperature of the boundary layer, and/or the presence of extended emission. The synthetic X-ray spectra consist of both absorbed and reflected components. In systems with massive, high-column density discs and viewing angles close to edge-on, the reflected continuum can dominate the X-ray emission. This effect is less pronounced in systems with low-mass, lower-column density discs. We explore {\it i}) systems going from $\delta$ to $\beta$ states, {\it ii}) $\delta$-types that become $\beta/\delta$ sources, {\it iii}) the variability of the three Fe emission lines in the 6.0--7.0 energy range, and {\it iv}) the possible physical processes behind the $\alpha$ sources. The observations from iconic symbiotic systems are discussed in line of the present models. Our framework offers predictive power for future X-ray monitoring and provides a path toward connecting accretion disc physics with observed spectral states in symbiotic binaries with accreting WDs.
\end{abstract}




\section{Introduction}
\label{sec:intro}

Symbiotic systems comprise of a white dwarf (WD) orbiting a late-type star, that can be a red giant or an asymptotic giant branch star \citep[see reviews by][]{Munari2019,Merc2025,Whitelock1987}. It is accepted that, as it orbits around its companion, the WD accretes from the wind of the late-type star. Although details might slightly change, simulations in the literature predict that such configuration produces an accretion disc around the WD \citep[see for example,][]{deValB2009,HuarteEspinosa2013,Lee2022,Liu2017,Mohamed2012,Nagae2004,Saladino2019}.

The presence of accretion discs in symbiotic systems may imprint different electromagnetic signatures. For example, it produces stochastic fluctuations in optical light curves known as flickering \citep{Dobrzycka1996,Gromadzki2013,Merc2024}. The double-peak emission line profiles associated with rotation of material has also been used to assess accretion disc sizes \citep[e.g.,][and references therein]{Robinson1994,Zamanov2024}, including those in the O\,~{\sc vi}~6825~\AA\, Raman line \citep{Lee1999,Lee2007}.

X-ray emission has also been used as a direct probe of accretion in symbiotic systems \citep{Mukai2017}. In this wavelength range, the presence of accretion disc is suggested by the 6.4 keV Fe fluorescent line in the X-ray spectra \citep[e.g.,][]{Ishida2009}. Nevertheless, only about 20 per cent of the Galactic symbiotic systems are known to have been  detected in X-rays \citep[][]{Merc2019}\footnote{See the statistics compiled by the New Online Database of Symbiotic Variables \url{https://sirrah.troja.mff.cuni.cz/~merc/nodsv/}}, and not all of them exhibit the 6.4 keV Fe line.

\citet{Murset1997} proposed a classification of X-ray-emitting symbiotic systems according to their spectral properties, later expanded by \citet{Luna2013}. $\alpha$-type symbiotic systems are those with super soft X-ray spectra ($E< 0.5$ keV), with their emission attributed to thermonuclear burning of H on the surface of the WD companion or jet activity \citep[e.g.,][]{Orio2007,Gonzalez-Riestra2013}. The $\beta$ class is assigned to those exhibiting spectra peaking at $\approx$0.8--1.0~keV. In general, they can be modelled with plasma at temperatures of a few times $10^{6}$~K and have been attributed to the presence of shocks produced by the winds of the companion. The $\delta$-type correspond to symbiotic systems with heavily-absorbed, hard ($kT > 5$ keV) plasma. The nature of the X-ray emission in these systems is attributed to the boundary layer, the region between the surface of the WD and the inner face of the accretion disc \citep[see][]{Pringle1979,Patterson1985}. They exhibit the presence of the 6.4 keV Fe emission line, in addition to the contribution of the He- and H-like Fe emission at 6.7 and 6.97 keV \citep{Eze2014}. Finally, $\beta/\delta$ objects share the properties of $\delta$-type systems but with an extra soft component.

However, high-quality X-ray observations keep pushing the limits and questioning the grounds on which the classification described above is built. Specially because some sources change their X-ray class from one to another. For example, the symbiotic recurrent nova system T CrB evolved from a $\delta$- into a $\beta/\delta$-type after entering a more active phase, as it approaches its next nova event \citep[][]{Luna2018,Toala2024_TCrB,Zhekov2019}. The symbiotic system RT Cru has been known to have a $\delta$-type spectrum  \citep{Luna2007,Danehkar2021}, but by 2019 it was reported that it changed for a few months to be more consistent with a $\beta$-type \citep[e.g.,][]{Pujol2023}. Another example is that of HM~Sge that was classified as a  $\beta$-type object in the original paper of \citet{Murset1997} using {\it ROSAT} data, but {\it XMM-Newton} observations obtained on 2016 discovered an extra super soft component very likely produced by its jet \citep{Toala2023a}.

The examples described above suggest that there should be more fundamental physics behind the extreme variations exhibited by X-ray-emitting symbiotic systems. For example, if one accepts that at least 80 per cent of them (if not all) have an accretion disc \citep[e.g.,][]{Merc2024}, one might even ask what is the impact of the accretion process in the X-ray emission. 

In \citet[][hereinafter Paper I]{Toala2024} we proposed that if all the X-ray emission comes from a boundary layer inside the accretion disc, the inclination angle and the presence of extended X-ray emission (hot bubbles and/or jets) can be use to explain the different X-ray types in the ($\alpha$, $\beta$, $\delta$, and $\beta/\delta$) classification scheme. In Paper I we used radiative transfer simulations on a simple disc geometry, a flared disc, to make our predictions and compare with observations. 

Paper I concluded that $\delta$-type symbiotic systems can be explained by systems seen near to an edge-on viewing angle (e.g., through the disc) for systems emitting hard X-ray emission. This configuration naturally produces a heavily-absorbed spectrum. Lower inclination systems (i.e., lower column density) resemble those of the $\beta$-type. $\beta/\delta$-type systems can only be  explained if the soft contribution has a smaller extinction and is outside the accretion disc, that is, if the emission corresponds to extended emission (a hot bubble and/or jets)  as seen in the well-characterised system R Aqr \citep{Kellogg2001,Nichols2007,Toala2022}. The $\alpha$-type was the most difficult to reconcile with the presence of an accretion disc, and was obtained in calculations with low-absorption column densities and face-on viewing angles.

In this paper we extend the work presented in Paper I by exploring synthetic X-ray emission obtained from more realistic density distributions. We performed hydrodynamic simulations that are subsequently used as input in radiative transfer simulations to  predict synthetic X-ray emission and the contribution from the reflecting material, the accretion disc. 

This paper is organized as follows. In Section~\ref{sec:methods} we present the details of our hydrodynamical simulations and the basic configuration of the radiative transfer calculations. In Section~\ref{sec:results} we present our results. A discussion, which involves the comparison with observations of variable X-ray symbiotic systems, is presented in Section~\ref{sec:discussion}. Finally, a summary and conclusions are listed in Section~\ref{sec:summary}.

\section{Methods}
\label{sec:methods}

\begin{figure}
\begin{center}
\includegraphics[width=\linewidth]{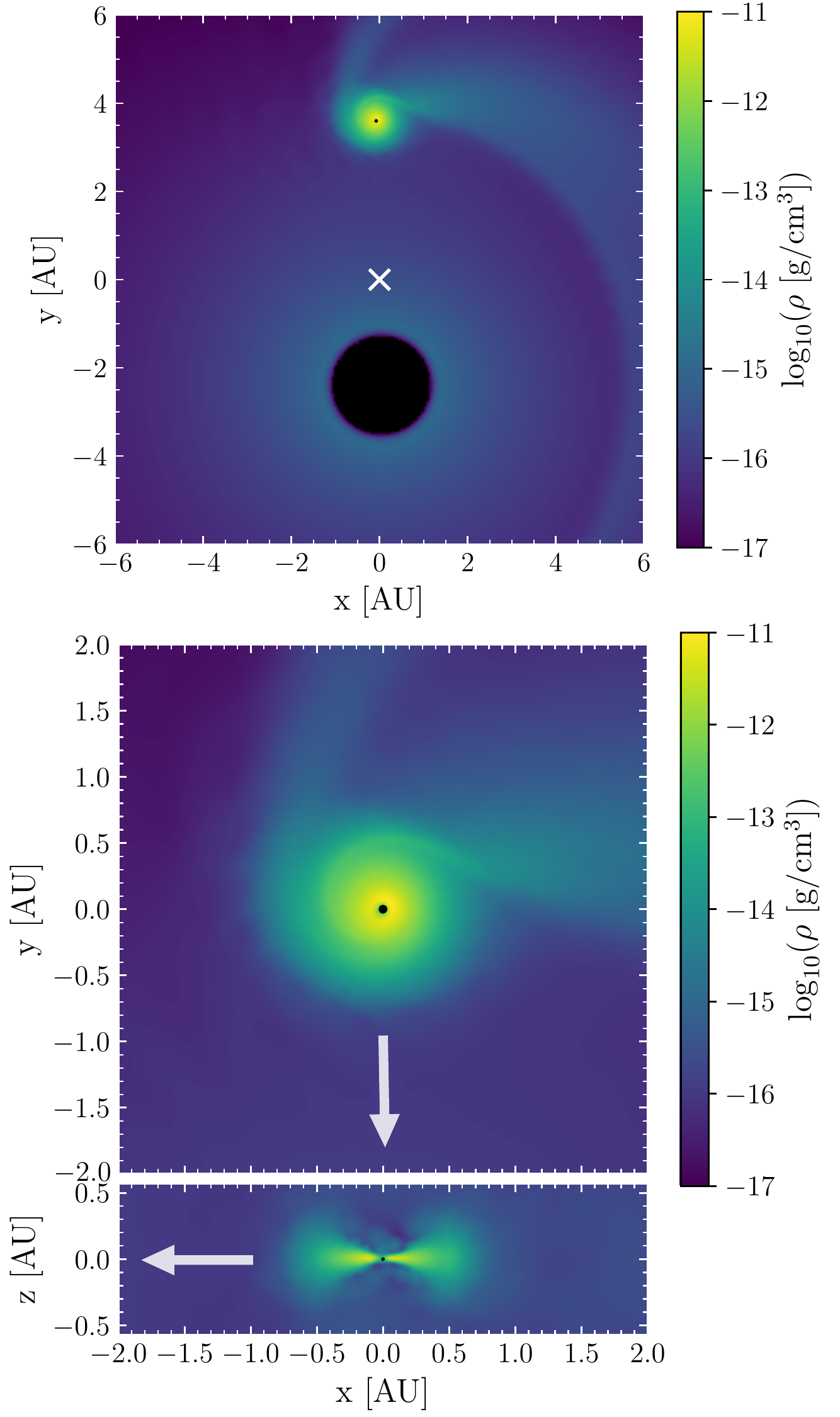}
\end{center}
\caption{Density maps of the {\sc phantom} simulation of a binary system in a circular orbit (see description in Section~\ref{sec:phantom}) at an arbitrary time. The top panel shows the binary components in the central $x-y$ plane, with the mass donor star shown with a black circle and the companion is surrounded by the accretion disc. The cross shows the position of the centre of mass. The bottom shows cuts in the $x-z$ and $x - y$ planes that correspond to  face-on and edge-on views of the companion to highlight the details of the accretion disc. The arrows point towards the direction of the mass donor star.}
\label{fig:den_maps}
\end{figure}

\subsection{Hydrodynamic simulations}
\label{sec:phantom}

We run smoothed-particle hydrodynamic (SPH) simulations with the {\sc phantom} code \citep{Price2018}. The numerical setup of the present simulations follows that described in \citet{Malfait2024}, which includes cooling by atomic hydrogen. The wind particles have an adiabatic equation of state defined by
\begin{equation}
    P = (\gamma - 1) \rho u,
    \label{eq:pressure}
\end{equation}
\noindent where $P$, $\rho$, $u$, and $\gamma$ are the pressure, gas density, internal energy, and the polytropic index, respectively. The gas temperature is then computed combining the ideal gas equation and Eq.~(\ref{eq:pressure}),
\begin{equation}
    P = \frac{\rho k T}{\mu m_\mathrm{H}},
\end{equation}
\noindent with $k$, $m_\mathrm{H}$, and $\mu$ as the Boltzmann's constant, the hydrogen mass, and the mean molecular weight. In the present simulations we adopt $\gamma=1.2$ and $\mu = 1.26$, values consistent for an atomic material with typical temperatures of AGB winds  \citep[see][]{Millar2004}.

\begin{figure*}
\begin{center}
\includegraphics[width=\linewidth]{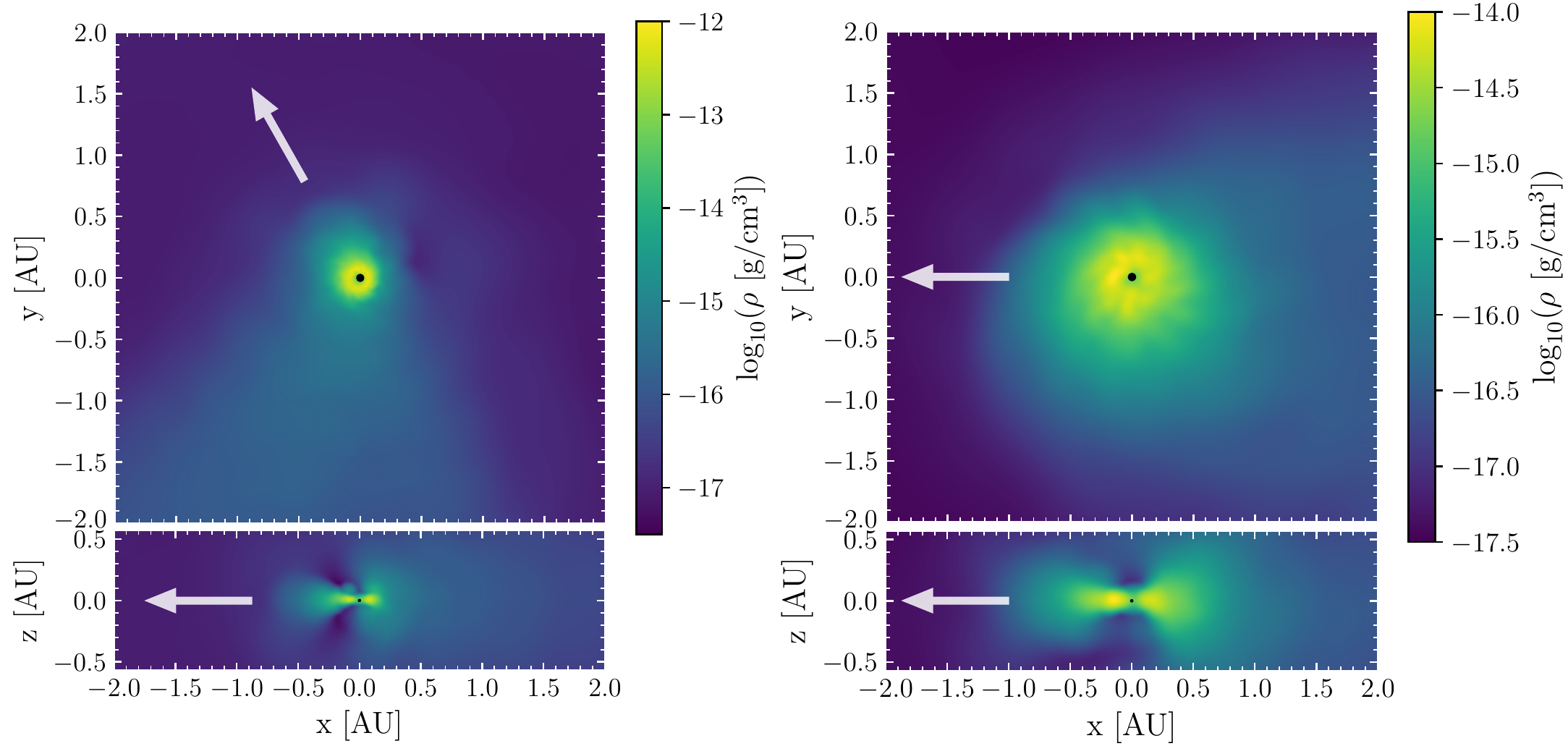}
\end{center}
\caption{Same as Fig.~\ref{fig:den_maps} but for the accretion disc during maximum of accretion (left; $t/P = 0.18$) and apastron passage (right; $t/P = 0.5$) in the eccentric ($e=0.45$) orbit simulation. Note the differences in the scale bars.}
\label{fig:raqr}
\end{figure*}

We start with a SPH simulation with almost exactly the same configuration as that presented in simulation v10e00 of \citet{Malfait2024}. This corresponds to a binary system in a circular orbit, where a giant mass losing star has a mass of $M_1 = 1.5$ M$_\odot$, a radius of $R_1=270$~R$_\odot$ and a luminosity of $L_1=5315$ L$_\odot$. The input wind velocity is 10 km~s$^{-1}$ with a constant mass-loss rate of $\dot{M}_\mathrm{w} = 10^{-7}$ M$_\odot$ yr$^{-1}$. The companion WD is modelled with a sink particle with a mass of $M_2 = 1$ M$_\odot$ and an accretion radius of $R_\mathrm{acc}=0.01$~AU. The system is initialized with a separation of $a=6$ AU, which results in an orbital period of $P=9.2$~yr.

The simulation runs until reaching a steady state, that is, until the mass accretion rate stalls around a constant value of $\dot{M}_\mathrm{acc} =10^{-8}$ M$_\odot$ yr$^{-1}$. At this point, the accretion disc around the sink particle has a size of $\lesssim0.7$ AU and has a flared morphology, very similar to the structures of the simulations presented in \citet[][]{Malfait2024}. The total estimated mass of the accretion disc, integrated in a spherical volume with radius of 1.5 AU and centred on the accreting WD is $8.9 \times 10^{-8}$~M$_\odot$\footnote{Similar accretion disc masses have been estimated by previous numerical works. For example, \citet{Lee2022} and \citet{Malfait2024} predict disc masses around [2--7]$\times10^{-8}$ M$_\odot$, while \citet{HuarteEspinosa2013} predicts disc mass of $\lesssim8\times10^{-6}$ M$_\odot$ for mass-loss rates of the donor star as high as $10^{-5}$~M$_\odot$ yr$^{-1}$.}. The density structure of the binary system and the accretion disc of this simulation are shown in Fig.~\ref{fig:den_maps}.

\begin{figure}
\begin{center}
\includegraphics[width=\linewidth]{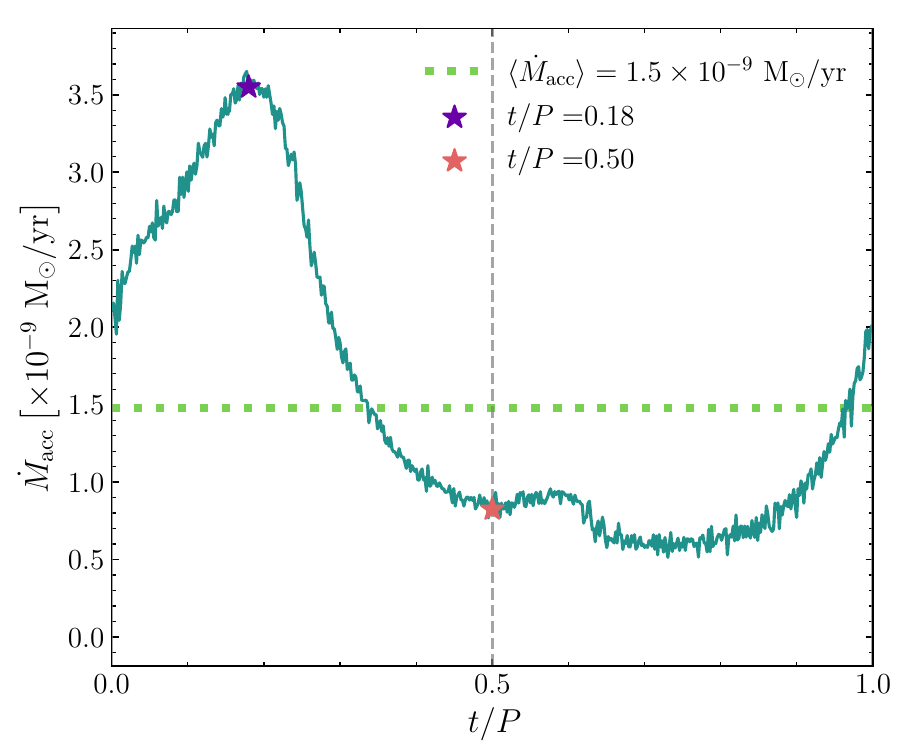}
\end{center}
\caption{Temporal evolution of the mass accretion rate $\dot{M}_\mathrm{acc}$ for the simulation with an eccentric system over one orbital period. $t/P=0.0$ represents the periastron position while $t/P=0.5$ indicates the system's apastron. The star symbols mark the moments of maximum accretion ($t/P=0.18$) and that of the apastron passage ($t/P=0.50$).}
\label{fig:mass_acc_his}
\end{figure}

A second simulation is performed taking into account an eccentric system with parameters close to those describing the symbiotic system R Aqr \citep[see][]{Alcolea2023},  that is, a primary mass of $M_1 = 1.0$ M$_\odot$, an accreting companion with a mass of $M_2 = 0.7$ M$_\odot$, an eccentricity of $e=0.45$, and orbital period of $P=42.2$ yr. Consequently, the semi-major axis resulted in $a=14.51$ AU. The mass-loss rate of the giant primary star is set to the same value as that used for the circular simulation, but the wind injection velocity is 12 km~s$^{-1}$.

As demonstrated by \citet{Malfait2024}, the properties of the accretion disc (mass and size) in simulations of eccentric binaries change within the orbit, but so does the mass accretion rate. In Fig.~\ref{fig:raqr} we show two snapshots of the simulations to illustrate the disc size during the maximum of accretion and the configuration during apastron passage. The masses of the accretion disc during these two specific times are $2.9 \times 10^{-9}$~M$_\odot$ and $1.6 \times 10^{-9}$~M$_\odot$, respectively.

The history of the mass accretion rate during one orbital period of the eccentric simulation is illustrated in Fig.~\ref{fig:mass_acc_his}, evolving around a median value of $\langle\dot{M}_\mathrm{acc}\rangle = 1.5\times10^{-9}$~M$_\odot$~yr$^{-1}$. The curve resembles the results of numerical simulations of eccentric cases where the peak of the accretion is not correlated with the periastron passage \citep[e.g.,][]{SaladinoPols2019}. Instead, the peak is shifted towards an orbital position just after periastron passage. Similarly, the minimum of accretion is not the apastron.

We note that using a sink particle to model the WD oversimplifies the accretion process. Adopting a sink particle hampers the proper formation of a boundary layer, which is the source of UV and X-ray emission in the accretion process. The presence of the boundary layer will be modelled with the aid of post-processing radiative transfer calculations described in the following subsection.

\subsection{Radiative transfer calculations}

\begin{figure}
\begin{center}
\includegraphics[width=\linewidth]{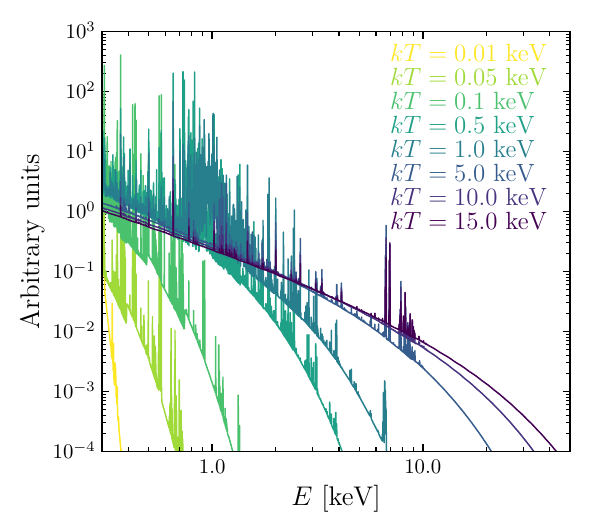}
\end{center}
\caption{Examples of \texttt{apec} model spectra in the 0.3--50.0 keV energy range used as input in the radiative transfer simulations.}
\label{fig:spec_apec}
\end{figure}

Radiative transfer calculations are performed with the code {\sc skirt} \citep[version 9.0;][]{Camps2020}. {\sc skirt} is a Monte Carlo radiative transfer code that models the effects of absorption, scattering, and re-emission from a source into a medium. This medium can be constructed with the aid of geometric structures or can be included from numerical simulations. In the present work we use the accretion disc density structures created in the SPH simulations described in Section~\ref{sec:phantom} and illustrated in Fig.~\ref{fig:den_maps} and \ref{fig:raqr}.

The code {\sc skirt} includes the treatment of X-ray photons by capturing the effects of Compton scattering on free electrons, photoabsorption, and fluorescence by cold atomic gas, scattering on bound electrons and extinction by dust \citep{VanderMeulen2023}. Although originally designed to study the X-ray properties of the central regions of active galactic nuclei, our team has exploited its capabilities to study accreting WD in symbiotic systems \citep[][]{Guerrero2025,VasquezTorres2024} and in the the context of Be/X-ray binaries \citep{Toala2025}.

Strictly speaking, the source of X-ray photons is the boundary layer, the region between the surface of the WD and the inner region of the accretion disc. However, we simplify the radiative transfer calculations by adopting the centre of the density structure (the position of the accreting WD) as the source of X-ray photons. We assume this source to be a thermal, collisionally-ionised plasma characterised with a temperature denoted as $kT$. For this, we used the \texttt{apec} model \citep{Smith2001} with abundances set to Solar \citep{Lodders2009}. We explored different plasma temperatures from 0.01 up to 50 keV. Some examples are illustrated in Fig.~\ref{fig:spec_apec}.

\begin{figure}
\begin{center}
\includegraphics[width=\linewidth]{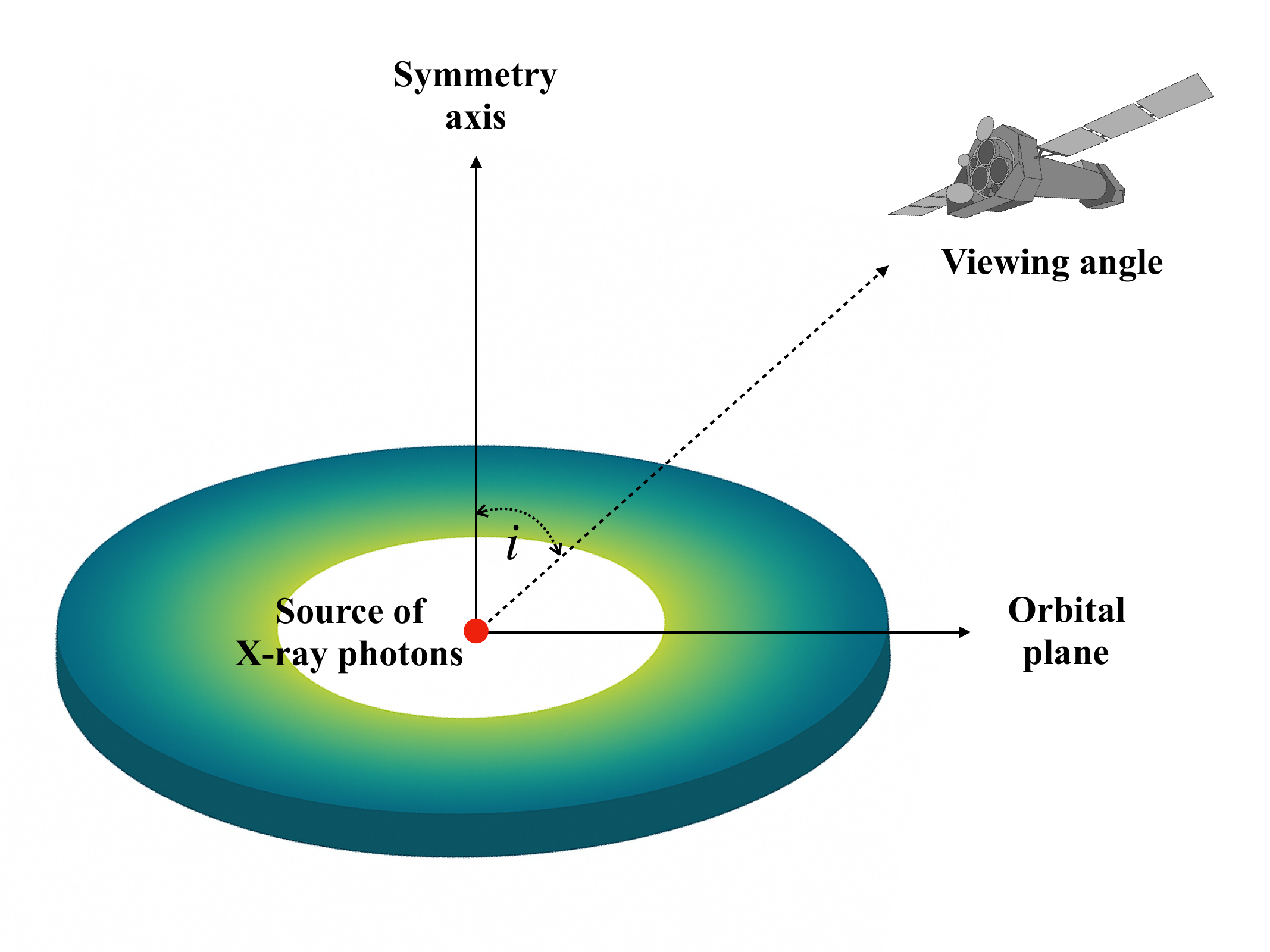}
\end{center}
\caption{A schematic representation of the source of X-ray photons inside the accretion disc. The inclination angle is denoted with $i$ and is measured from the symmetry axis.}
\label{fig:disc}
\end{figure}

{\sc skirt} is able to consistently compute the absorption and the reflection\footnote{Including scattering and fluorescent processes.} properties of the X-ray spectra for specific viewing angle configurations. The viewing angle $i$ will be defined as sketched in Fig.~\ref{fig:disc}, with $i=0^\circ$ corresponding to a face on view, while $i=90^\circ$ is defined as the edge-on configuration.

All calculations are performed adopting an X-ray luminosity of 10~L$_\odot$ in the 0.3--10 keV energy range for the source of X-ray photons and a distance of 200 pc, unless other values are specified.

\section{Results}
\label{sec:results}

\begin{figure*}
\begin{center}
\includegraphics[width=0.33\linewidth]{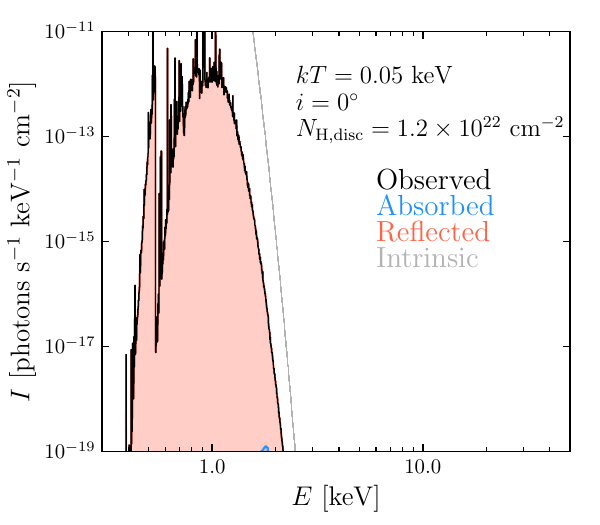}~
\includegraphics[width=0.33\linewidth]{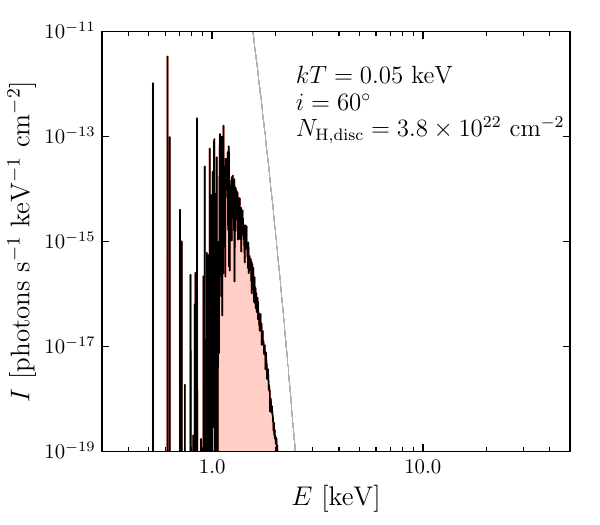}~
\includegraphics[width=0.33\linewidth]{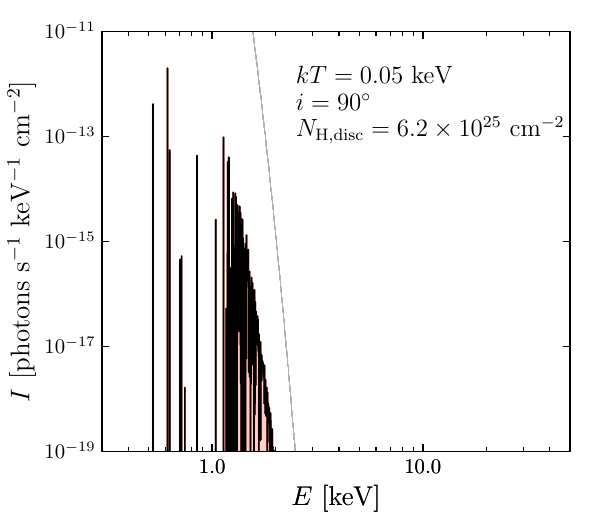}\\
\includegraphics[width=0.33\linewidth]{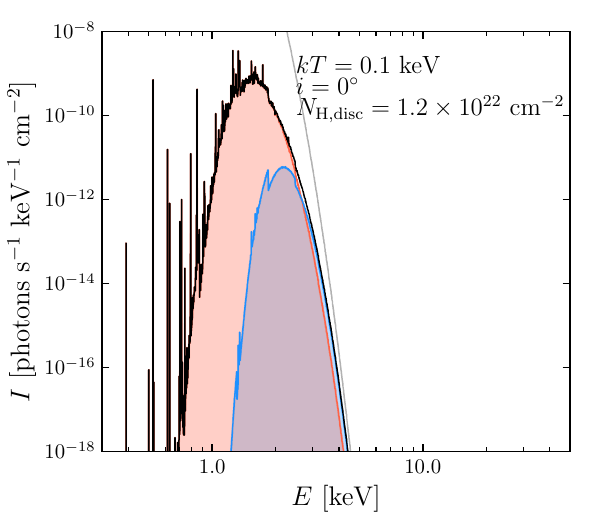}~
\includegraphics[width=0.33\linewidth]{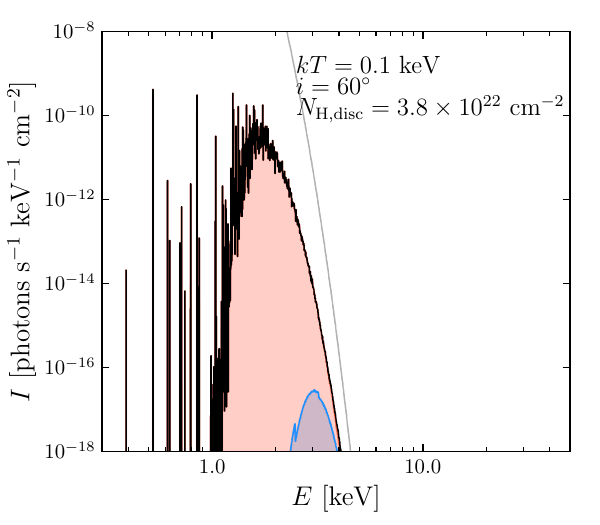}~
\includegraphics[width=0.33\linewidth]{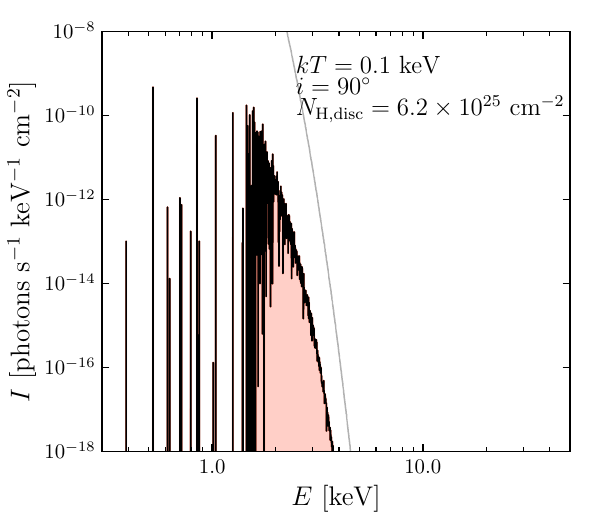}\\
\includegraphics[width=0.33\linewidth]{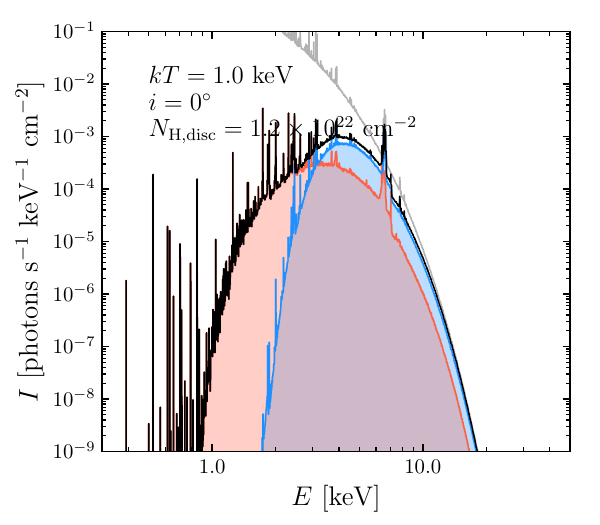}~
\includegraphics[width=0.33\linewidth]{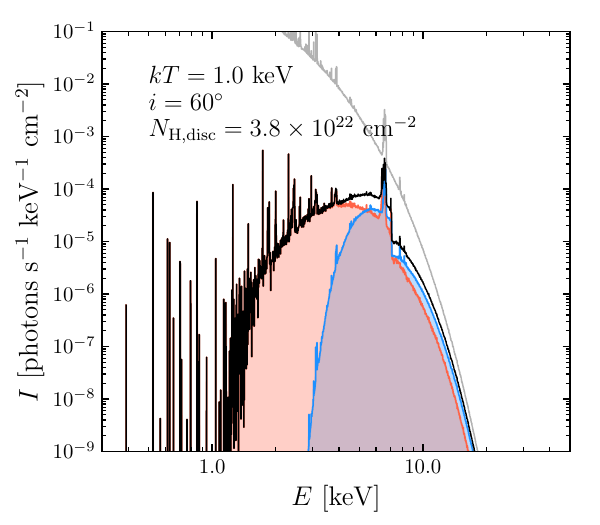}~
\includegraphics[width=0.33\linewidth]{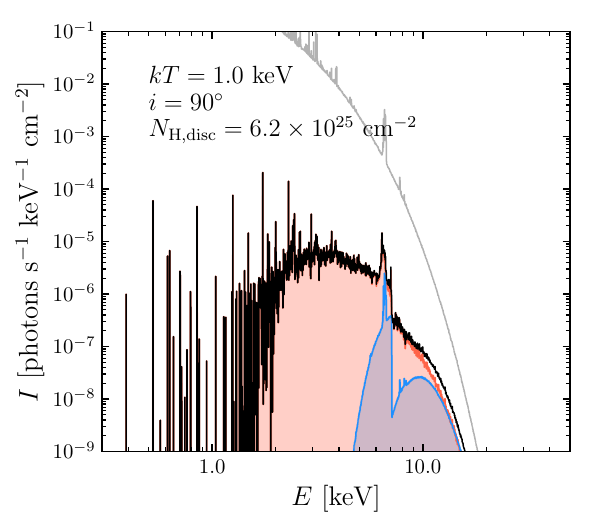}\\
\includegraphics[width=0.33\linewidth]{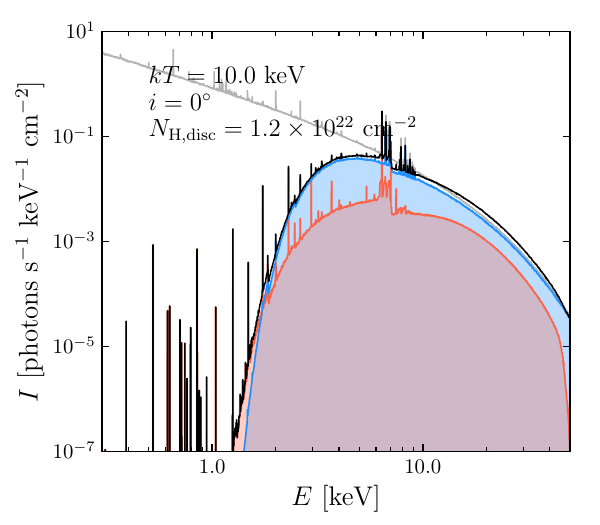}~
\includegraphics[width=0.33\linewidth]{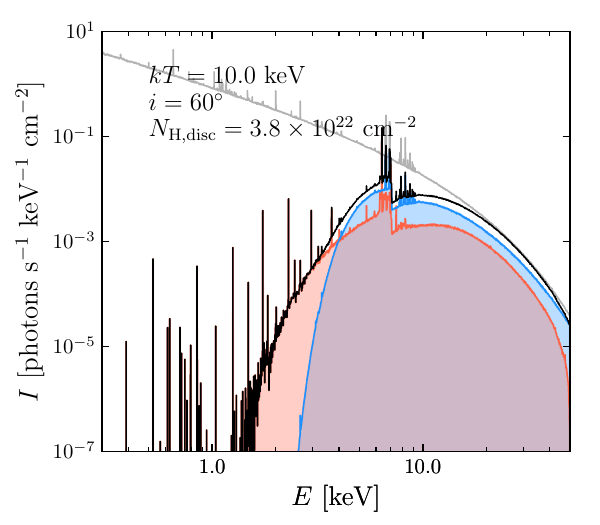}~
\includegraphics[width=0.33\linewidth]{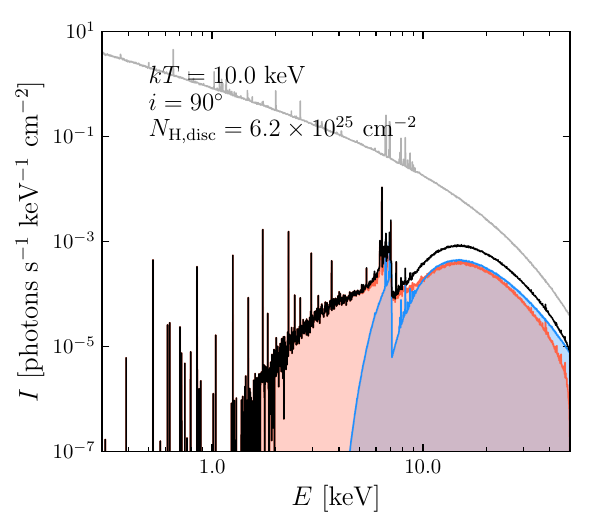}
\end{center}
\caption{Synthetic X-ray spectra obtained from {\sc skirt} radiative transfer simulations applied to the accretion disc resulted from the circular orbit case (Fig.~\ref{fig:den_maps}). Different panels show the components of the different synthetic spectra of simulations with plasma temperatures of $kT=0.05$ (top row), 0.1 (second row), 1.0 (third row), and 10.0 keV (bottom row). The left, central, and right columns show the results from simulations with viewing angles of $i=0$, 60, and 90$^\circ$, respectively. Note the different scales between the panes in the top row.}
\label{fig:comp}
\end{figure*}

\begin{figure*}
\begin{center}
\includegraphics[width=0.33\linewidth]{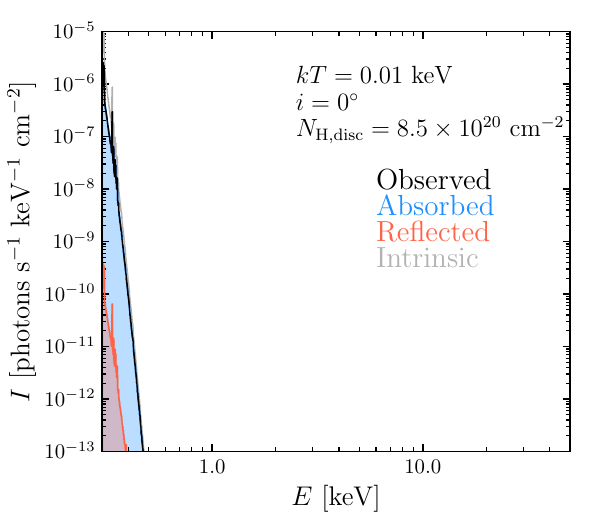}~
\includegraphics[width=0.33\linewidth]{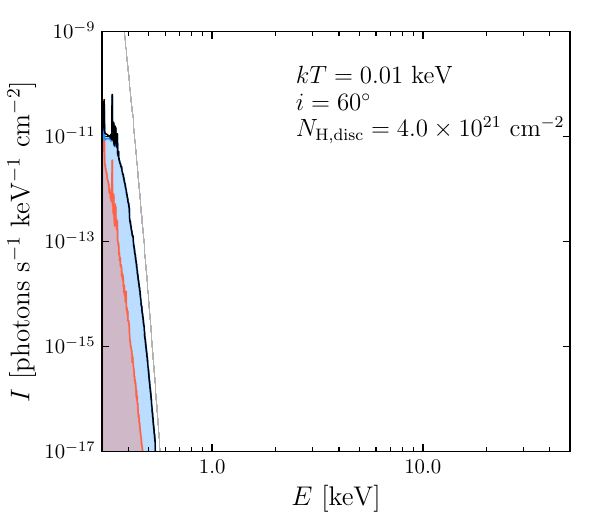}~
\includegraphics[width=0.33\linewidth]{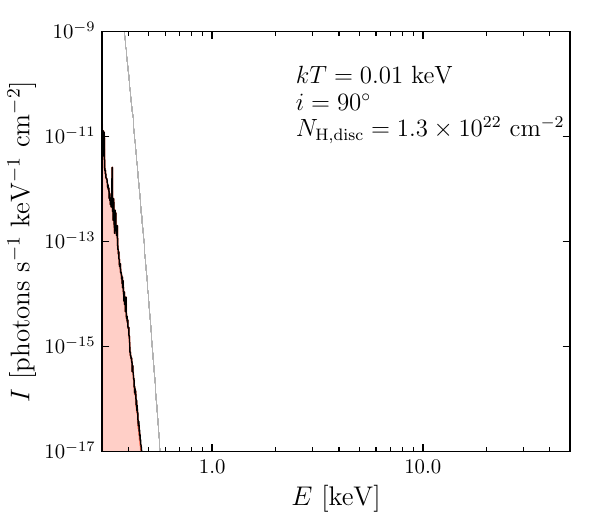}\\
\includegraphics[width=0.33\linewidth]{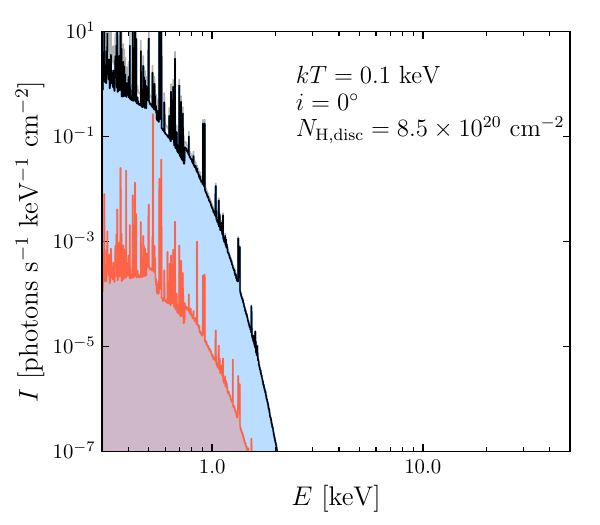}~
\includegraphics[width=0.33\linewidth]{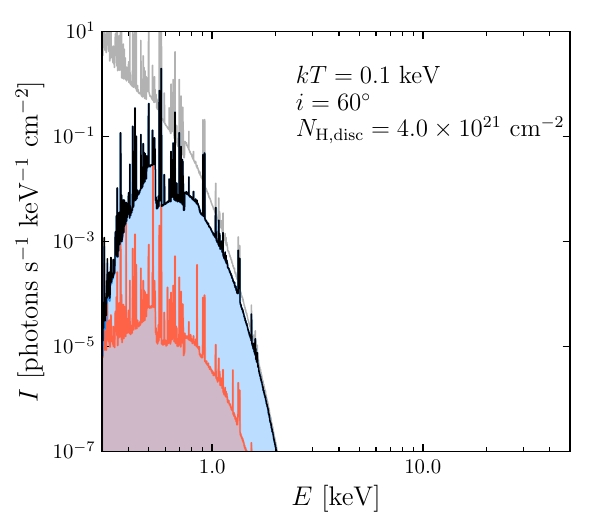}~
\includegraphics[width=0.33\linewidth]{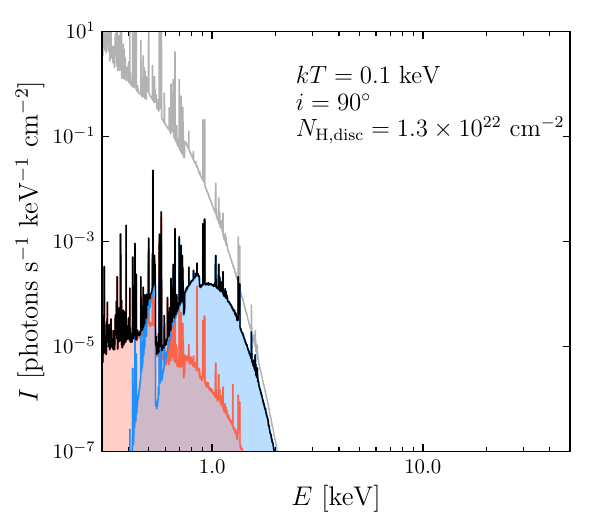}\\
\includegraphics[width=0.33\linewidth]{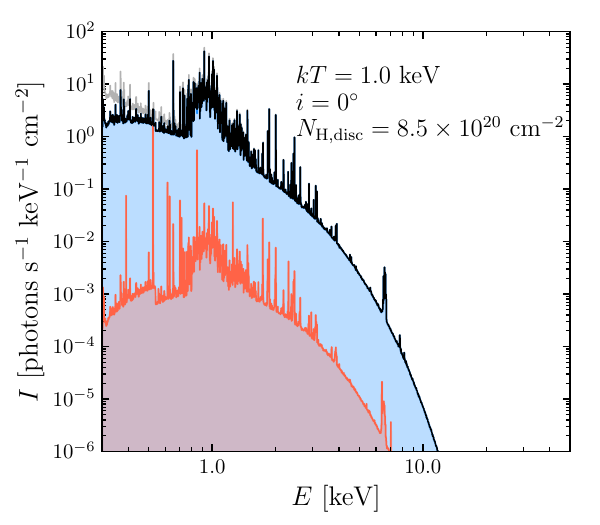}~
\includegraphics[width=0.33\linewidth]{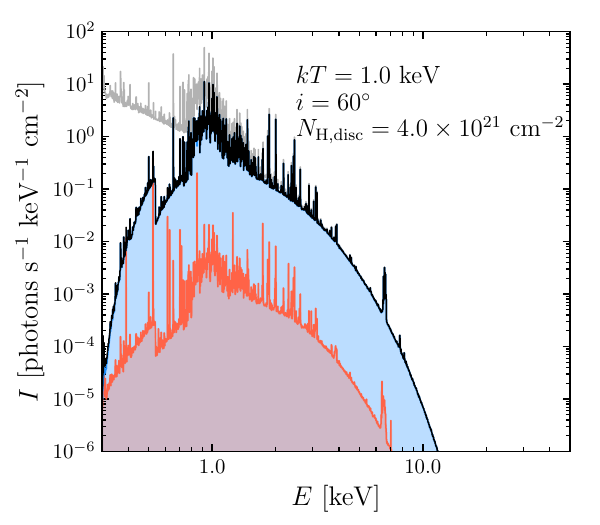}~
\includegraphics[width=0.33\linewidth]{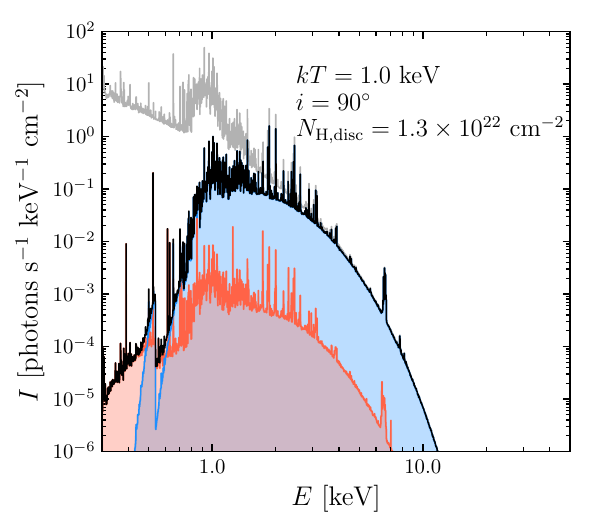}\\
\includegraphics[width=0.33\linewidth]{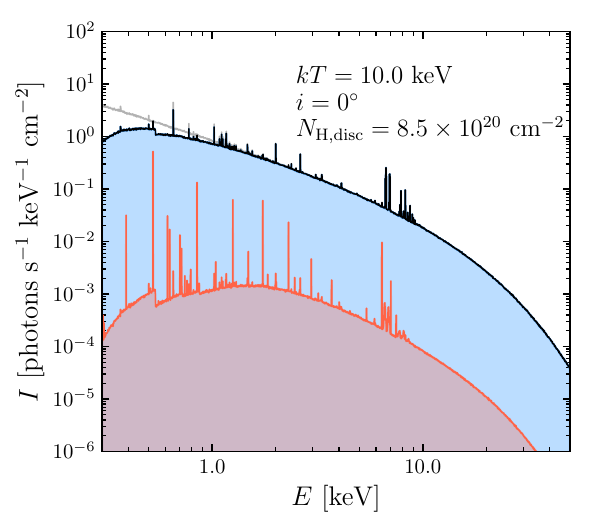}~
\includegraphics[width=0.33\linewidth]{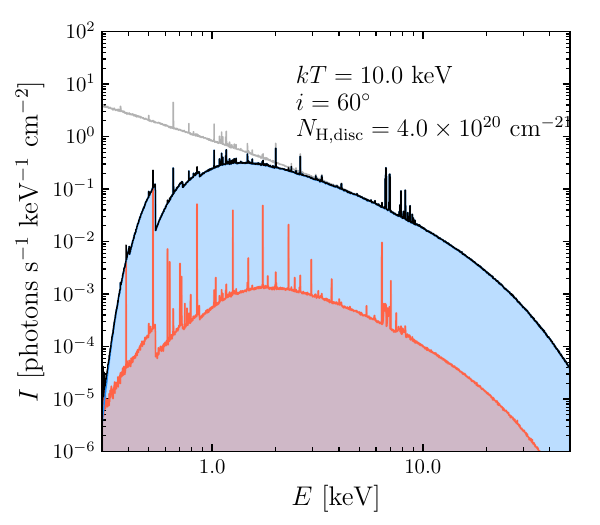}~
\includegraphics[width=0.33\linewidth]{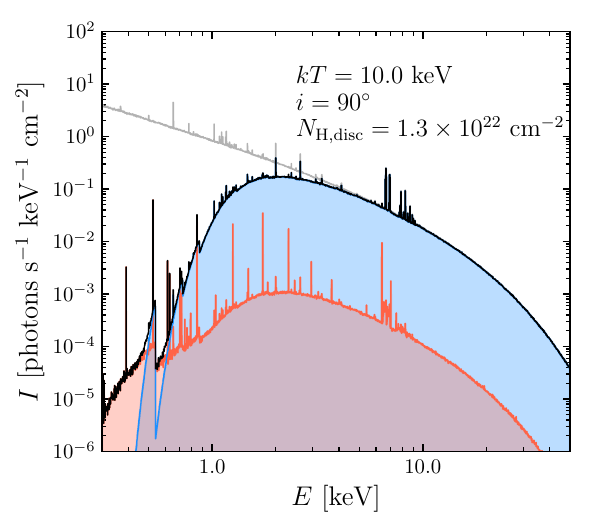}
\end{center}
\caption{Same as Fig.~\ref{fig:comp} but for {\sc skirt} simulations obtained from the accretion disc in the eccentric orbit going through the apastron passage (Fig.~\ref{fig:raqr} right panel).}
\label{fig:comp_min}
\end{figure*}

Fig.~\ref{fig:comp} shows examples of X-ray spectra obtained by applying {\sc skirt} to the disc structure obtained from the circular orbit simulation. Different panels show the results of adopting different plasma temperatures for the boundary layer, in combination with different viewing angles. For comparison, Fig.~\ref{fig:comp_min} presents the results obtained using the accretion disc from the eccentric orbit simulation during apastron passage ($t/P = 0.5$; Fig.~\ref{fig:raqr} right panel).

The synthetic spectra presented in Fig.~\ref{fig:comp} and \ref{fig:comp_min} help confirming previous findings that the iron fluorescent line at 6.4 keV becomes prominent for simulations with $kT \geq 1.0$ keV. The He- and H-like Fe lines at 6.7 and 6.97 keV, respectively, also become important in simulations with plasma temperatures $kT \geq 5$ keV. This situation seems to hold regardless of the adopted viewing angle $i$ (see bottom panels in those figures).

\begin{figure}
\begin{center}
\includegraphics[width=\linewidth]{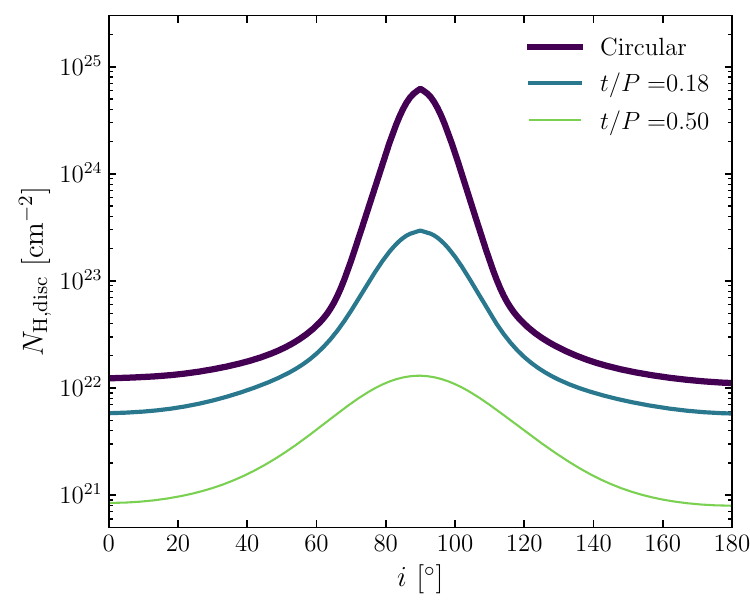}
\end{center}
\caption{Averaged column density profile integrated along different line of sights $i$ through the density distribution of the accretion disc of Fig.~\ref{fig:den_maps}. Different lines show the results from the circular orbit, the eccentric case at the maximum of accretion ($t/P=0.18$) and that obtained during apastron passage ($t/P = 0.50$).}
\label{fig:den_col}
\end{figure}

As expected, the spectra obtained by integrating along the accretion disc (e.g., $i$ values close to an edge-on view) are the ones experiencing the highest absorption. It is important to note that the extinction in the spectra presented in Fig.~\ref{fig:comp} and \ref{fig:comp_min} is only attributed to material in the accretion disc itself, which naturally varies depending on the viewing angle $i$. To illustrate this effect, we plot in Fig.~\ref{fig:den_col} the net column density produced by our disc models ($N_\mathrm{H,disc}$) as a function of the inclination angle.

Fig.~\ref{fig:den_col} shows that for the case of the accretion disc obtained from the circular orbit simulation, $N_\mathrm{H,disc}$ is larger than $10^{23}$ cm$^{-2}$ for viewing angles close to edge-on view ($i \gtrsim 70^\circ$), very similar to that resulted from spectral analysis of $\delta$ and $\beta/\delta$ sources \citep[][]{Stute2009,Luna2013,Toala2024_TCrB,VasquezTorres2024}, with a peak  at $N_\mathrm{H,disc}=6.1\times10^{24}$ cm$^{-2}$ at $i=90^\circ$. 
The distribution has a basal value of $N_\mathrm{H,disc}=1.2\times10^{22}$~cm$^{-2}$ for a face on viewing angle ($i=0^\circ$). Which mean that even for small inclination angles ($i\approx 0$) the column density is significant. This effect is clearly seen in all panels of the left column of Fig.~\ref{fig:comp}, where the soft X-ray emission ($E< 0.8$ keV) is heavily-absorbed. In fact, this figure does not present the results obtained for $kT=0.01$ keV because their resultant fluxes are insignificant.

In contrast, the column density distribution of the accretion disc obtained at the apastron passage in the eccentric case ($t/P = 0.50$) more than an order of magnitude smaller than that of the circular case for $i\approx 0^\circ$, but almost three orders of magnitude of difference for $i \approx 90^\circ$ (Fig.~\ref{fig:den_col}). Naturally, the synthetic X-ray spectra obtained using this accretion disc model experience smaller extinction, allowing the production of synthetic spectra with a considerable soft X-ray component (see Fig.~\ref{fig:comp_min}).

Fig.~\ref{fig:comp} and \ref{fig:comp_min} also illustrates the contribution from the two main components to the synthetic observed spectra obtained with {\sc skirt}. Each panel shows the absorbed spectra and the component produced by reflection in the accretion disc. Most of the spectra presented in Fig.~\ref{fig:comp} are dominated by the reflecting component as a consequence of the high column density produced by the accretion disc. On the other hand, the contribution of the reflecting component in the synthetic X-ray spectra of Fig.~\ref{fig:comp_min} is generally less important.

The radiative transfer simulations presented here allow us to produce the $\alpha$-, $\beta$-, and $\delta$-type X-ray spectra of symbiotic systems. Similarly to what we reported in Paper I, the X-ray spectra of the $\delta$-type can only be achieved for boundary layers with high plasma temperatures seen through the accretion disc, that is, close to edge-on viewing angles (large $i$). The extremely high column densities obtained from the analyses of $\delta$-type symbiotic systems are produced by the accretion disc.

The $\beta$-type can be reproduced by different combinations of plasma temperature and viewing angle. According to Fig.~\ref{fig:comp} and \ref{fig:comp_min}, $\beta$-type spectra can be achieved by boundary layers with plasma temperatures $kT \approx 1$ keV with no strong restriction on the viewing angle. If the boundary layer has higher plasma temperature, $\beta$-type spectra can also be produced by viewing angles close to face-on view, small $i$ values (see bottom left panel in Fig.~\ref{fig:comp_min}).

We predict that the formation of $\alpha$-type X-ray spectra takes place in calculations of low-density accretion discs, in low inclination viewing angles. The top panels of Fig.~\ref{fig:comp_min} show that for $N_\mathrm{H} > 10^{21}$ cm$^{-2}$, this emission will be heavily extinguished, making it difficult to be detected.

\section{Discussion} 
\label{sec:discussion}

\begin{figure*}
\begin{center}
\includegraphics[width=0.5\linewidth]{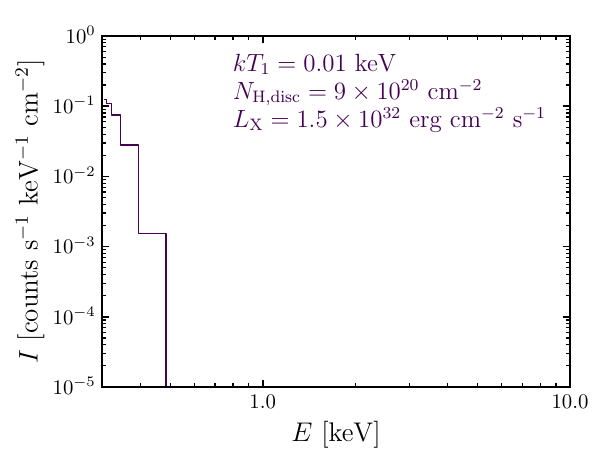}~
\includegraphics[width=0.5\linewidth]{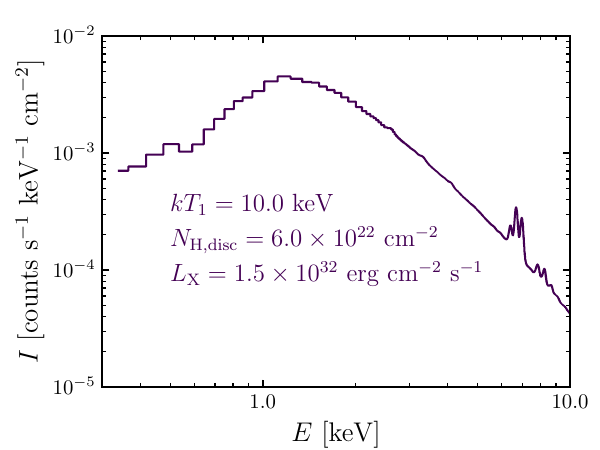}\\
\includegraphics[width=0.5\linewidth]{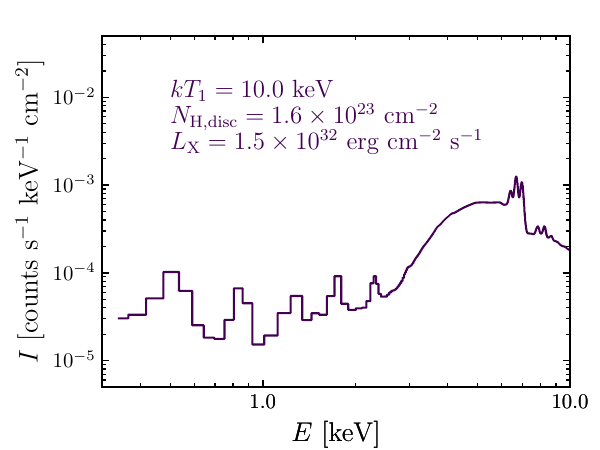}~
\includegraphics[width=0.5\linewidth]{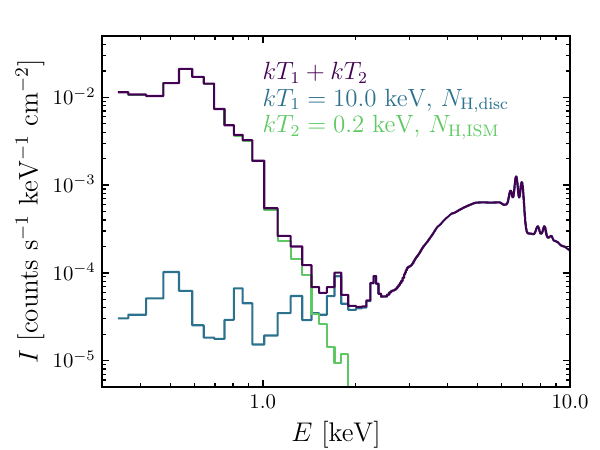}
\end{center}
\caption{Examples of synthetic spectra for $\alpha$ (top left), $\beta$ (top right), $\delta$ (bottom left), and $\beta/\delta$ (bottom right) systems convolved with {\it XMM-Newton} EPIC-pn camera with the medium optical filter response matrix. The  $\beta/\delta$ spectrum was created combining the $\delta$ model from the bottom left panel with a soft plasma component $kT_2$ absorbed by an ISM column density of $N_\mathrm{H,ISM}=10^{21}$ cm$^{-2}$. All spectra were binned requesting a minimum of 20 counts per energy bin.} 
\label{fig:epic_abd}
\end{figure*}

\begin{figure}
\begin{center}
\includegraphics[width=\linewidth]{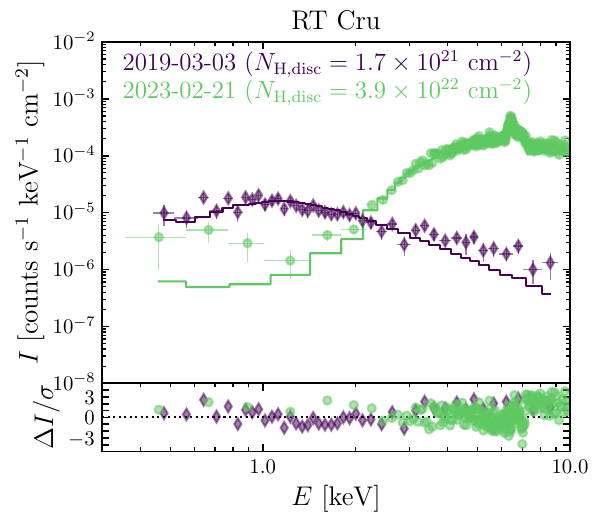}
\end{center}
\caption{Background-subtracted {\it XMM-Newton} EPIC-pn spectra of the symbiotic system RT Cru. Different symbols represent two different epochs. Both spectra were modelled using the same plasma temperature  50 keV, but different $N_\mathrm{H,disc}$. The solid line histograms are the best fits to the data using models from the present paper (see Section~\ref{sec:b_to_bd}). The bottom panel shows the residuals of the fits.}
\label{fig:RT_Cru}
\end{figure}

In the following subsections we address the observed behaviour in the X-ray spectra of iconic symbiotic systems using our synthetic data. The analysis was performed making use of the X-Ray Spectral Fitting Package \citep[{\sc xspec}, version 12.12.1;][]{Arnaud1996}. We converted our synthetic spectra into additive one-parameter fit tables with the help of the {\sc heasoft}\footnote{\url{https://heasarc.gsfc.nasa.gov/docs/software/heasoft/}} task \texttt{ftflx2tab}\footnote{\url{https://heasarc.gsfc.nasa.gov/lheasoft/help/ftflx2tab.html}} to make them readable in {\sc xspec}.

The synthetic spectra are then convolved with the {\it XMM-Newton} EPIC-pn calibration matrices\footnote{Those corresponding to a medium optical blocking filter. For further details see \url{https://xmm-tools.cosmos.esa.int/external/xmm\_user\_support/documentation/uhb/epicfilters.html}}  for a direct comparison with available X-ray data. In Fig.~\ref{fig:epic_abd} we show examples of $\alpha$-, $\beta$-, and $\delta$-type spectra which were obtained by adopting different plasma temperatures for the boundary layers, $kT$, and different contributing column density from their accretion discs, $N_\mathrm{H,disc}$. In all cases the models were normalised to an X-ray luminosity of $L_\mathrm{X} = 1.5\times10^{32}$~erg~cm$^{-2}$~s$^{-1}$ in the 0.3--10.0 keV energy range, which seems to be a representative value for all X-ray emitting symbiotic systems \citep[see figure 5 in][]{Guerrero2024}. The synthetic EPIC-pn spectra of Fig.~\ref{fig:epic_abd} were binned requesting a minimum of 20 counts per energy bin.

The X-ray data discussed in the following sections were retrieved from the XMM-Newton Science Archive\footnote{\url{https://nxsa.esac.esa.int/nxsa-web/\#search}}. Their analysis and reduction followed standard procedures within the Science Analysis System \citep[{\sc sas}, version 20.0;][]{Gabriel2004}.  Whenever necessary, spectral analysis was performed using {\sc xspec}, where the best fit to the data was assessed with the reduced $\chi^2$ statistics ($\chi^{2}_\mathrm{DoF}$).

\subsection{From $\beta$- to $\delta$-type}
\label{sec:b_to_bd}

We begin by highlighting a notable feature between the $\beta$-type (top-right panel) and $\delta$-type (bottom-left panel) spectra of Fig.~\ref{fig:epic_abd}. Both were computed assuming the same boundary layer plasma temperature, $kT = 10$ keV. The only difference lies in the net column density (i.e., mass) of the accretion disc. The $\beta$ spectrum was generated using the accretion disc from the eccentric simulation at its peak accretion ($t/P=0.18$; Fig.~\ref{fig:raqr}, left panel), whereas the $\delta$ spectrum corresponds to the more massive disc obtained in the circular orbit SPH simulation (Fig.~\ref{fig:den_maps}). The net column densities differ by a factor of $\approx 2.7$, which alone is sufficient to explain transitions in X-ray-emitting symbiotic systems from $\beta$- to $\delta$-type spectra without changes in boundary-layer temperature. An iconic symbiotic system that exhibited such behaviour is RT Cru.

The first detection of high-energy photons from RT Cru was obtained with {\it INTEGRAL} \citep{Chernyakova2005, Sguera2012, Sguera2015}. This source can be classified as $\delta$-type, with the boundary layer exhibiting a maximum shocked temperature of $\approx$53 keV and a mass accretion rate of a few times $10^{-9}$~M$_\odot$~yr$^{-1}$ \citep{Luna2007,Luna2018_RT_Cru}. The multi-epoch, multi-mission observations analysed by \citet{Danehkar2021} and \citet{Danehkar2024} revealed X-ray variability on both short and long timescales. In particular, the variability observed in the soft band appears to be driven either by changes in the net column density or by the intrinsic variability of the boundary layer. \citet{Pujol2023} reported that, over several months in 2019, RT Cru transitioned to a $\beta$-type spectrum, accompanied by a reduction in mass accretion rate by two orders of magnitude, while maintaining the same plasma temperature of $\approx$53 keV.

As an experiment, we used our synthetic spectra computed for $kT = 50$~keV to perform spectral modelling of the publicly available {\it XMM-Newton} EPIC-pn observations of RT Cru. These correspond to data obtained on 2019-03-03 (Obs. ID 0831790801; PI: N.~Schartel) and 2023-02-20 (Obs. ID 0920750101; PI: F.~Walter). The best-fit model to the 2019 observation, which resulted in $\chi^{2}_\mathrm{DoF} = 1.76$, yields an absorption-corrected X-ray flux of $F_\mathrm{X} = 2.3\times10^{-13}$erg~cm$^{-2}$s$^{-1}$, corresponding to an X-ray luminosity of $L_\mathrm{X} = 1.7\times10^{32}$erg~s$^{-1}$ at a distance of 2.52 kpc \citep{BailerJones2021}, in close agreement with the value reported by \citet{Pujol2023}. For the 2023 observation, we derive a flux of $F_\mathrm{X} = 3.0\times10^{-11}$erg~cm$^{-2}$~s$^{-1}$, equivalent to a luminosity of $L_\mathrm{X} = 2.3\times10^{34}$erg~s$^{-1}$. The model has a quality of $\chi^2_\mathrm{DoF} = 1.39$.

A comparison between our models and the observations is shown in Fig.\ref{fig:RT_Cru}. While the match is not perfect -- likely due to the specific density structures resulted from our simulations, not tailored for RT Cru -- it nevertheless  confirms that the reduction in accretion rate during 2019 is linked to a decrease in both the mass and structural complexity of the accretion disc. This, in turn, affects the X-ray emission and produces the observed $\beta$-type spectrum. Furthermore, our analysis indicates that in the most recent observation, RT Cru has reached flux levels comparable to those reported in 2005 and 2009 \citep[see Table 1 in][]{Pujol2023}, corresponding to an estimated mass accretion rate\footnote{We adopt a WD mass of 1.2 M$_\odot$ \citep{Luna2007} and a standard radius of 0.01 R$_\odot$.} of about $\dot{M}_\mathrm{acc} \approx 4\times10^{-9}$~M$_\odot$~yr$^{-1}$.

This analysis also implies that not all $\beta$-type sources are systems in which the X-ray-emitting plasma is produced by wind collisions within the symbiotic system, as suggested in the literature \citep[e.g.,][]{Merc2019}. Therefore, we suggest caution when interpreting $\beta$-type spectra.

\subsection{From $\delta$- to $\beta/\delta$-type}

Similar to what was found in Paper I, our radiative-transfer simulations presented here show that $\delta$-type spectra can be obtained with models where the boundary layer has a high plasma temperature ($kT > 5$ keV) and when the accretion disc imprints a high column density to the line of sight ($N_\mathrm{H,disc}$). The later condition is achieved by dense (massive) accretion discs and/or viewing angles close to an edge-on configuration.

There are some cases in the literature that show a transformation from a $\delta$-type source into a $\beta/\delta$-type by producing a new, considerable softer, spectral component ($kT_2 \approx 0.2$ keV). Spectral modelling of $\beta/\delta$-type sources have shown that the softer component cannot be fitted by the same high absorption as that of the hard component \citep[e.g.,][]{Stute2009,Zhekov2019,Luna2018}. It requires a smaller column density with values similar to those of the ISM ($N_\mathrm{H,ISM}$). Such unavoidable result suggest that the X-ray-emitting gas producing the soft component can not be located within the accretion disc, close to the accreting WD. That is, the soft plasma cannot be related to the boundary layer (if the $\delta$ component of the spectrum is still present). Consequently, we advice against using the contribution to the X-ray luminosity of the soft component, because it can overestimate the true mass accretion rate.

\citet{Lucy2020} argued that the newly developed soft spectral component of the $\beta/\delta$ source V694~Mon was best attributed to the development of bipolar outflows, just outside the accretion disc. These authors suggested that the outflow helps stabilizing the accretion disc during a period of high accretion rate. A similar situation was proposed by \citet{Toala2024_TCrB}, where soft X-ray emission appeared in the symbiotic recurrent nova system T CrB correlated with the onset of a period of high activity. High-dispersion X-ray data of the soft X-ray emission from T CrB revealed emission lines, supporting the idea that the soft X-ray emission is produced by shocks -- very likely due to a bipolar outflow with expansion velocity $>100$ km s$^{-1}$ -- instead of being associated with a blackbody-like emitter.

The super soft X-ray components in the $\beta/\delta$ sources CH Cyg and R Aqr have been spatially-resolved by {\it Chandra} \citep{Karovska2010,Sacchi2024}, and they are also attributed to the presence of jets with velocities a few times 100 km~s$^{-1}$. \citet{Toala2022} argued that the extended super soft hot bubbles detected in R Aqr with {\it XMM-Newton} is yet another evidence for jet activity in this system. They argued that the extended hot bubbles were created by the action of the jets through the so-called jet feedback mechanism \citep[see][and references therein]{Soker2016}.

The bottom-right panel of Fig.~\ref{fig:RT_Cru} presents an example of a $\beta/\delta$ synthetic X-ray spectrum. We took the $\delta$-type spectrum of the bottom-left panel of this figure, in combination with a soft apec model ($kT_2=0.2$ keV) absorbed by a typical $N_\mathrm{H,ISM} = 10^{21}$~cm$^{-2}$. In contrast, the $\delta$ component has a column density imprinted by the accretion disc of $N_\mathrm{H,disc}=1.6\times10^{23}$~cm$^{-2}$. A similar column density applied to the soft component will completely extinguish it.

The presence and variability of jets in symbiotic systems are not the only factors driving changes in the X-ray spectra of $\beta/\delta$ sources. Recent multi-epoch {\it Chandra} and {\it XMM-Newton} analyses of R~Aqr by \citet{Sacchi2024} and \citet{VasquezTorres2024} revealed that variations in the hard X-ray emission reflect changes in the mass accretion process. Observations spanning 22 years showed that the luminosity of the boundary layer increases, peaking shortly after periastron passage. This behaviour is consistent with the modified Bondi–Hoyle–Lyttleton accretion model proposed by \citet{VasquezTorres2024} \citep[see also][]{TejedaToala2025}, as well as with the eccentric orbit simulations presented in the current work.

\subsection{Variable Fe emission lines}

One of the most extensively studied X-ray-emitting symbiotic systems is CH~Cyg. It has been observed by nearly all major X-ray observatories and its spectrum has been consistently detected as a $\beta/\delta$-type \citep{Leahy1995, Leahy1987, Ezuka1998, Karovska2007, Mukai2007, Toala2023b}. Owing to the exceptionally high quality of its spectra and its resemblance to Seyfert galaxies, \citet{Wheatley2006} applied analysis techniques commonly used for X-ray-emitting AGN and argued that the observed spectrum is likely dominated by scattering.

\begin{figure}
\begin{center}
\includegraphics[width=\linewidth]{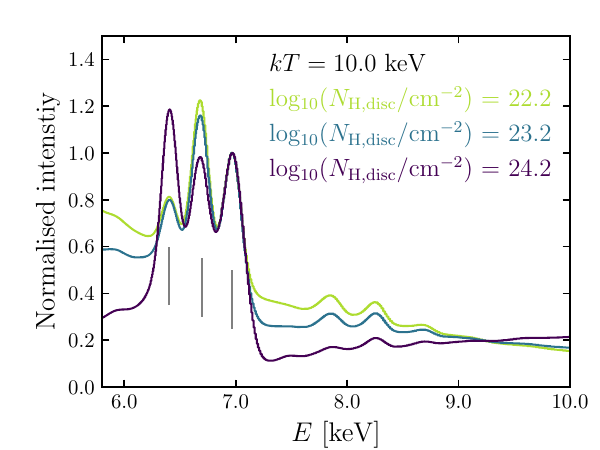}
\end{center}
\caption{Normalised synthetic {\it XMM-Newton} EPIC-pn spectra of models with $kT = 10$ keV but integrated through different column densities of the accretion disc from the circular simulation. The position of the fluorescent, He-like, and H-like Fe emission lines at 6.4, 6.7, and 6.97 keV, respectively, are shown with vertical line segments.}
\label{fig:epic_Fe}
\end{figure}

The multi-epoch X-ray analysis presented by \citet{Mukai2007} provided insight into the variability of the Fe emission lines in CH Cyg. They showed that during the 1994 {\it ASCA} and 2001 {\it Chandra} observations, the He-like line at 6.7 keV dominated over the other two components. By 2006, however, {\it Suzaku} observations revealed that the fluorescent 6.4 keV line had become dominant. Subsequent analysis by \citet{Toala2023b} indicated that in the 2009 {\it Chandra} data the 6.7 keV line once again prevailed, while in the 2018 {\it XMM-Newton} observations the dominance of the fluorescent line was restored.

The prevalence of the 6.4 keV fluorescent line appears to be correlated with CH Cyg’s orbital period, estimated at about 15~yr \citep[see, e.g.,][and references therein]{Hinkle2009}. This suggests that the alternating dominance between the lines may be explained by periodic structural changes in the accretion disc. 

In Fig.~\ref{fig:epic_Fe} we demonstrate that if the plasma temperature of the boundary layer is keep constant, but one changes the properties of the accretion disc, we can explain the ratios of the three Fe emission lines. The 6.4 keV fluorescent line dominates over the other two when the column density (the mass of the disc) increases. 

CH Cyg is an eccentric symbiotic system, and it is natural to expect its accretion disc to undergo variations similar to those seen in the eccentric simulations. However, these variations are likely to be less pronounced, given that CH Cyg’s eccentricity ($e=0.12$) is significantly smaller than the value adopted in our simulations in Section~\ref{sec:phantom} ($e=0.45$). A detailed analysis of the X-ray properties of CH Cyg by means of a tailored reflection models is being prepared (Vasquez-Torres et al. in prep.).

\subsection{On the super soft $\alpha$ sources}

\begin{figure}
\begin{center}
\includegraphics[width=\linewidth]{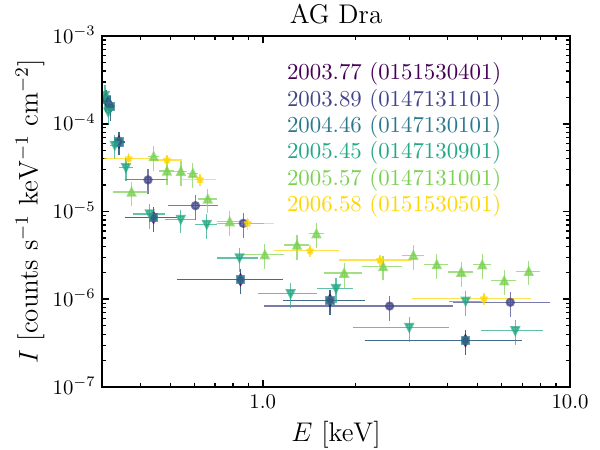}
\end{center}
\caption{Background-subtracted EPIC-pn spectra of AG Dra. Different symbols show spectra from different epochs. The numbers in parenthesis are the Obs. ID. of the data.}
\label{fig:AG_Dra}
\end{figure}

According to our results, $\alpha$ sources should be observable primarily through low column density sightliness and/or in a face-on orientation with respect to the system's orbit. Given their intrinsically soft spectra ($kT \sim 0.01$ keV), any increase in column density -- whether caused by variable ejections from the mass-donor companion or structural changes in the accretion disc -- could effectively extinguish this emission.

Their super soft spectra ($E < 0.5$ keV) has been attributed to the tail of the stellar continuum of the accreting WD companion in the system spilling into the soft X-ray range \citep{Jordan1994,Murset1997}. 
For example, \citet{Gonzalez-Riestra2013} presented the analysis of {\it XMM-Newton} data of RR Tel in combination with non-LTE stellar models and found that the super-soft component seems to be consistent with stellar models with temperatures $T_\mathrm{eff} \approx 150$~kK.

However, the spectral resolution in the super soft range provided by the current generation X-ray instruments is not sufficient enough to discriminate from a shocked plasma with temperatures of 0.01 keV (see top panel of Fig.~\ref{fig:comp_min}). When the X-ray emission is consistent with jets observed at other wavelengths it should provide strong evidence against the interpretation based solely on stellar continuum from the WD. For example, the case of HM~Sge discussed in \citet{Toala2023a}, where the appearance of the super-soft component in the 2016 {\it XMM-Newton} observations with plasma temperature of $kT=0.03$ keV ($=3.5\times10^{5}$~K) is consistent with the $\sim100$~km~s$^{-1}$ jet velocity detected in optical wavelengths \citep{Corradi1999}.

Unfortunately, there are not many $\alpha$ sources reported in the literature\footnote{For example, according to the New Online Database of Symbiotic Variables, only four sources in our Galaxy exhibit an $\alpha$-type spectrum: AG~Dra, DD Mic, RR Tel, and StHA 32. See \url{https://sirrah.troja.mff.cuni.cz/~merc/nodsv/utilities/x-rays.html}}, but the most observed is the iconic AG Dra. Fig.~\ref{fig:AG_Dra} compiles all available {\it XMM-Newton} EPIC-pn spectra of this $\alpha$ source, showing that the super-soft component is absent in the 2005.57 and 2006.58 epochs. 

\citet{Skopal2015} used the data from 2005.45 epoch -- which shows a strong contribution from the super-soft component -- while disregarding the other available datasets, and derived an effective temperature for the accreting WD of $T_\mathrm{eff} \approx 160{,}000$~K.
In contrast, \citet{Sion2012} argued against temperatures exceeding $T_\mathrm{eff} = 80{,}000$ K, suggesting instead that the super-soft X-ray emission in AG Dra may originate from a different source, such as a boundary layer.

However, we note that the clear presence of harder X-ray emission in the spectra of AG Dra (see Fig.\ref{fig:AG_Dra}) may represent the true signature of the boundary layer. In contrast, the temperature associated with the intermittent super-soft emission ($kT \approx 0.01$ keV) appears consistent with the series of bipolar ejections observed in AG Dra, with velocities of $\sim$70–200 km s$^{-1}$ \citep{Torbett1987,Mija2002,Ogley2002,Lee2019}.

A comprehensive analysis of $\alpha$-type symbiotic systems including all available observations, particularly for AG~Dra, is needed to further understand the production of super soft emission in these systems.

\section{Summary and conclusions}
\label{sec:summary}

We studied the impact of the accretion disc in the X-ray emission from symbiotic systems by means of SPH simulations in combination with post-processing radiative-transfer calculations. This study represents an improvement over that of Paper I, where we adopted geometric density structures modelled with a simple flared disc.

We ran SPH simulations with the code {\sc phanton} to model binary systems in circular and eccentric ($e=0.45$) orbits. The density structures produced by the accretion discs were then introduced into the radiative-transfer code {\sc skirt} to make predictions on their synthetic X-ray spectra. The source of X-ray photons in accreting symbiotic systems is the boundary layer, but for simplicity in our calculations is adopted to be a point-source at the position of the WD. The input plasma temperature was varied from 0.01 to 50 keV. 

We corroborate the findings of Paper I, which proposed a unified scenario for X-ray-emitting symbiotic systems. In this framework, the observed X-ray properties depend on a combination of factors: the presence and characteristics of the accretion disc, the system’s viewing angle, the plasma temperature of the boundary layer, and the possible contribution from jets and/or hot bubbles. In the present work, we extend this picture by examining how variations in these components can drive the observed diversity and temporal variability of X-ray emission in symbiotic systems. By bridging accretion disc physics with observed spectral states, our framework not only advances the interpretation of current data but also provides predictive capability for future X-ray monitoring of symbiotic binaries.

Our specific findings are:

\begin{itemize}

\item The observed X-ray emission is affected by a combination of the density structure of the accretion disc, the viewing angle, and the plasma temperature of the boundary layer. By exploring all these parameters we are able to reproduce the X-ray spectra of symbiotic systems of the $\alpha$-, $\beta$-, and $\delta$-type.

\item The synthetic X-ray spectra consist of both absorbed and reflected components. In systems with massive, high-column density discs and viewing angles close to edge-on, the reflected continuum can dominate the X-ray emission. This effect is less pronounced in systems with low-mass, lower-column density discs.

\item Reflection features in the synthetic X-ray spectra, such as the Fe K$\alpha$ 6.4 keV line, are sensitive to the geometry and optical depth of the accretion disc. This line's contribution becomes significant only for boundary layers with high plasma temperatures ($kT \gtrsim 5$ keV).

\item We explore spectral changes in systems transitioning between $\delta$- and $\beta$-type states using models with a fixed plasma temperature for the boundary layer but variable accretion disc properties. Focusing on the case of RT~Cru, we conclude that during its low mass accretion phase, the disc reduced significantly its mass, producing the observed spectral changes. Analysis of the 2023 {\it XMM-Newton} EPIC-pn observation suggests that the mass accretion rate has since recovered to values comparable to those estimated prior to the reduction ($\approx4\times10^{-9}$~M$_\odot$~yr$^{-1}$).

\item $\beta/\delta$ sources can only be explained if the extinction affecting the $\beta$ component is caused by a relatively small column density, comparable to that provided by the ISM. Consistent with the findings of Paper I, we attribute the transition of sources from the $\delta$ to the $\beta/\delta$ type to the development of bipolar ejections of material. Thus, we advise against using the luminosity of the $\beta$ component when estimating the mass accretion rate from $\beta/\delta$-type symbiotic systems.

\item The variability of the fluorescent, He-like, and H-like Fe emission lines at 6.4, 6.7, and 6.97 keV, respectively, can be attributed to changes in the accretion disc, assuming a fixed plasma temperature for the boundary layer. In the specific case of CH Cyg, these variations are likely driven by density changes in the accretion disc induced by the eccentric orbit of its accreting WD.

\item In agreement with Paper I, we find that $\alpha$-type sources arise from low plasma temperatures combined with either low column densities or inclination angles close to a face-on view of the system. However, the true nature behind the production of such soft plasma is debatable. Further multi-epoch analysis of super soft systems is most needed.

\end{itemize}

\section*{Acknowledgements} 

The authors thank an anonymous referee for comments and suggestions that improved the presentation and discussion of the original manuscript. J.A.T. thanks the staff of Facultad de Ciencias de la Tierra y el Espacio of Universidad Aut\'{o}noma de Sinaloa (FACITE-UAS, Mexico) for their support during a research visit. J.A.T. and D.A.V.T. acknowledge support from the UNAM PAPIIT project IN102324. D.A.V.T. thanks Secretaria de Ciencia, Humanidades, Tecnolog\'{i}a e Innovaci\'{o}n (SECIHTI, Mexico) for a student grant. This work is based on observations obtained with {\it XMM-Newton}, an European Science Agency (ESA) science mission with instruments and contributions directly funded by ESA Member States and NASA. This work has made a large use of NASA’s Astrophysics Data System (ADS).

\section*{DATA AVAILABILITY}

The X-ray data underlying this article were retrieved from the public archive of {\it XMM-Newton}. The hydrodynamic and radiative-transfer simulations underlying this article will be shared on reasonable request to the corresponding author.



\begin{thebibliography}{99}

\bibitem[Alcolea et al.(2023)]{Alcolea2023} Alcolea, J., Mikolajewska, J., G{\'o}mez-Garrido, M., et al.\ 2023, Highlights on Spanish Astrophysics XI, 190. 

\bibitem[Arnaud(1996)]{Arnaud1996} Arnaud, K.~A.\ 1996, Astronomical Data Analysis Software and Systems V, 101, 17. 

\bibitem[Bailer-Jones et al.(2021)]{BailerJones2021} Bailer-Jones, C.~A.~L., Rybizki, J., Fouesneau, M., et al.\ 2021, \aj, 161, 3, 147

\bibitem[Camps \& Baes(2020)]{Camps2020} Camps, P. \& Baes, M.\ 2020, Astronomy and Computing, 31, 100381

\bibitem[Chernyakova et al.(2005)]{Chernyakova2005} Chernyakova, M., Courvoisier, T.~J.-L., Rodriguez, J., et al.\ 2005, The Astronomer's Telegram, 519, 1. 

\bibitem[Corradi et al.(1999)]{Corradi1999} Corradi, R.~L.~M., Ferrer, O.~E., Schwarz, H.~E., et al.\ 1999, \aap, 348, 978

\bibitem[Danehkar et al.(2024)]{Danehkar2024} Danehkar, A., Drake, J.~J., \& Luna, G.~J.~M.\ 2024, \apj, 972, 1, 109

\bibitem[Danehkar et al.(2021)]{Danehkar2021} Danehkar, A., Karovska, M., Drake, J.~J., et al.\ 2021, \mnras, 500, 4, 4801

\bibitem[de Val-Borro et al.(2009)]{deValB2009} de Val-Borro, M., Karovska, M., \& Sasselov, D.\ 2009, \apj, 700, 2, 1148

\bibitem[Dobrzycka et al.(1996)]{Dobrzycka1996} Dobrzycka, D., Kenyon, S.~J., \& Milone, A.~A.~E.\ 1996, \aj, 111, 414


\bibitem[Eze(2014)]{Eze2014} Eze, R.~N.~C.\ 2014, \mnras, 437, 1, 857

\bibitem[Ezuka et al.(1998)]{Ezuka1998} Ezuka, H., Ishida, M., \& Makino, F.\ 1998, \apj, 499, 1, 388

\bibitem[Gabriel et al.(2004)]{Gabriel2004} Gabriel, C., Denby, M., Fyfe, D.~J., et al.\ 2004, Astronomical Data Analysis Software and Systems (ADASS) XIII, 314, 759. 

\bibitem[Gonz{\'a}lez-Riestra et al.(2013)]{Gonzalez-Riestra2013} Gonz{\'a}lez-Riestra, R., Selvelli, P., \& Cassatella, A.\ 2013, \aap, 556, A85

\bibitem[Gromadzki et al.(2013)]{Gromadzki2013} Gromadzki, M., Miko{\l}ajewska, J., \& Soszy{\'n}ski, I.\ 2013, \actaa, 63, 4, 405

\bibitem[Guerrero et al.(2025)]{Guerrero2025} Guerrero, M.~A., Vasquez-Torres, D.~A., Rodr{\'\i}guez-Gonz{\'a}lez, J.~B., et al.\ 2025, \aap, 693, A203

\bibitem[Guerrero et al.(2024)]{Guerrero2024} Guerrero, M.~A., Montez, R., Ortiz, R., et al.\ 2024, \aap, 689, A62

\bibitem[Hinkle et al.(2009)]{Hinkle2009} Hinkle, K.~H., Fekel, F.~C., \& Joyce, R.~R.\ 2009, \apj, 692, 2, 1360

\bibitem[Huarte-Espinosa et al.(2013)]{HuarteEspinosa2013} Huarte-Espinosa, M., Carroll-Nellenback, J., Nordhaus, J., et al.\ 2013, \mnras, 433, 1, 295

\bibitem[Ishida et al.(2009)]{Ishida2009} Ishida, M., Okada, S., Hayashi, T., et al.\ 2009, \pasj, 61, S77

\bibitem[Jordan et al.(1994)]{Jordan1994} Jordan, S., Murset, U., \& Werner, K.\ 1994, \aap, 283, 475

\bibitem[Karovska et al.(2010)]{Karovska2010} Karovska, M., Gaetz, T.~J., Carilli, C.~L., et al.\ 2010, \apjl, 710, 2, L132

\bibitem[Karovska et al.(2007)]{Karovska2007} Karovska, M., Carilli, C.~L., Raymond, J.~C., et al.\ 2007, \apj, 661, 2, 1048

\bibitem[Kellogg et al.(2001)]{Kellogg2001} Kellogg, E., Pedelty, J.~A., \& Lyon, R.~G.\ 2001, \apjl, 563, 2, L151

\bibitem[Leahy \& Volk(1995)]{Leahy1995} Leahy, D.~A. \& Volk, K.\ 1995, \apj, 440, 847

\bibitem[Leahy \& Taylor(1987)]{Leahy1987} Leahy, D.~A. \& Taylor, A.~R.\ 1987, \aap, 176, 262. 

\bibitem[Lee et al.(2019)]{Lee2019} Lee, Y.-M., Lee, H.-W., Lee, H.-G., et al.\ 2019, \mnras, 487, 2, 2166

\bibitem[Lee et al.(2022)]{Lee2022} Lee, Y.-M., Kim, H., \& Lee, H.-W.\ 2022, \apj, 931, 2, 142

\bibitem[Lee \& Kang(2007)]{Lee2007} Lee, H.-W. \& Kang, S.\ 2007, \apj, 669, 2, 1156

\bibitem[Lee \& Park(1999)]{Lee1999} Lee, H.-W. \& Park, M.-G.\ 1999, \apjl, 515, 2, L89

\bibitem[Liu et al.(2017)]{Liu2017} Liu, Z.-W., Stancliffe, R.~J., Abate, C., et al.\ 2017, \apj, 846, 2, 117

\bibitem[Lodders et al.(2009)]{Lodders2009} Lodders, K., Palme, H., \& Gail, H.-P.\ 2009, Landolt B{\"o}rnstein, 4B, 712

\bibitem[Lucy et al.(2020)]{Lucy2020} Lucy, A.~B., Sokoloski, J.~L., Munari, U., et al.\ 2020, \mnras, 492, 3, 3107

\bibitem[Luna et al.(2018a)]{Luna2018} Luna, G.~J.~M., Mukai, K., Sokoloski, J.~L., et al.\ 2018a, \aap, 619, A61

\bibitem[Luna et al.(2018b)]{Luna2018_RT_Cru} Luna, G.~J.~M., Mukai, K., Sokoloski, J.~L., et al.\ 2018b, \aap, 616, A53

\bibitem[Luna et al.(2013)]{Luna2013} Luna, G.~J.~M., Sokoloski, J.~L., Mukai, K., et al.\ 2013, \aap, 559, A6

\bibitem[Luna \& Sokoloski(2007)]{Luna2007} Luna, G.~J.~M. \& Sokoloski, J.~L.\ 2007, \apj, 671, 1, 741

\bibitem[Malfait et al.(2024)]{Malfait2024} Malfait, J., Siess, L., Esseldeurs, M., et al.\ 2024, \aap, 691, A84

\bibitem[Merc(2025)]{Merc2025} Merc, J.\ 2025, Galaxies, 13, 3, 49

\bibitem[Merc et al.(2024)]{Merc2024} Merc, J., Beck, P.~G., Mathur, S., et al.\ 2024, \aap, 683, A84

\bibitem[Merc et al.(2019)]{Merc2019} Merc, J., G{\'a}lis, R., \& Wolf, M.\ 2019, Research Notes of the American Astronomical Society, 3, 2, 28


\bibitem[Miko{\l}ajewska(2002)]{Mija2002} Miko{\l}ajewska, J.\ 2002, \mnras, 335, 1, L33

\bibitem[Millar(2004)]{Millar2004} Millar, T.~J.\ 2004, Asymptotic Giant Branch Stars, 247

\bibitem[Mohamed \& Podsiadlowski(2012)]{Mohamed2012} Mohamed, S. \& Podsiadlowski, P.\ 2012, Baltic Astronomy, 21, 88

\bibitem[Mukai et al.(2007)]{Mukai2007} Mukai, K., Ishida, M., Kilbourne, C., et al.\ 2007, \pasj, 59, 177

\bibitem[Munari(2019)]{Munari2019} Munari, U.\ 2019, , arXiv:1909.01389

\bibitem[M\"{u}rset et al.(1997)]{Murset1997} Muerset, U., Wolff, B., \& Jordan, S.\ 1997, \aap, 319, 201. 

\bibitem[Mukai(2017)]{Mukai2017} Mukai, K.\ 2017, \pasp, 129, 976, 062001

\bibitem[Nagae et al.(2004)]{Nagae2004} Nagae, T., Oka, K., Matsuda, T., et al.\ 2004, \aap, 419, 335

\bibitem[Nichols et al.(2007)]{Nichols2007} Nichols, J.~S., DePasquale, J., Kellogg, E., et al.\ 2007, \apj, 660, 1, 651

\bibitem[Ogley et al.(2002)]{Ogley2002} Ogley, R.~N., Chaty, S., Crocker, M., et al.\ 2002, \mnras, 330, 4, 772

\bibitem[Orio et al.(2007)]{Orio2007} Orio, M., Zezas, A., Munari, U., et al.\ 2007, \apj, 661, 2, 1105

\bibitem[Patterson \& Raymond(1985)]{Patterson1985} Patterson, J. \& Raymond, J.~C.\ 1985, \apj, 292, 535

\bibitem[Price et al.(2018)]{Price2018} Price, D.~J., Wurster, J., Tricco, T.~S., et al.\ 2018, \pasa, 35, e03

\bibitem[Pringle \& Savonije(1979)]{Pringle1979} Pringle, J.~E. \& Savonije, G.~J.\ 1979, \mnras, 187, 777

\bibitem[Pujol et al.(2023)]{Pujol2023} Pujol, A., Luna, G.~J.~M., Mukai, K., et al.\ 2023, \aap, 670, A32

\bibitem[Robinson et al.(1994)]{Robinson1994} Robinson, K., Bode, M.~F., Skopal, A., et al.\ 1994, \mnras, 269, 1

\bibitem[Sacchi et al.(2024)]{Sacchi2024} Sacchi, A., Karovska, M., Raymond, J., et al.\ 2024, \apj, 961, 1, 12

\bibitem[Saladino \& Pols(2019)]{SaladinoPols2019} Saladino, M.~I. \& Pols, O.~R.\ 2019, \aap, 629, A103

\bibitem[Saladino et al.(2019)]{Saladino2019} Saladino, M.~I., Pols, O.~R., \& Abate, C.\ 2019, \aap, 626, A68

\bibitem[Sguera et al.(2015)]{Sguera2015} Sguera, V., Bird, A.~J., \& Sidoli, L.\ 2015, The Astronomer's Telegram, 8448, 1. 

\bibitem[Sguera et al.(2012)]{Sguera2012} Sguera, V., Drave, S.~P., Sidoli, L., et al.\ 2012, The Astronomer's Telegram, 3887, 1. 

\bibitem[Sion et al.(2012)]{Sion2012} Sion, E.~M., Moreno, J., Godon, P., et al.\ 2012, \aj, 144, 6, 171

\bibitem[Skopal(2015)]{Skopal2015} Skopal, A.\ 2015, \na, 36, 116

\bibitem[Smith et al.(2001)]{Smith2001} Smith, R.~K., Brickhouse, N.~S., Liedahl, D.~A., et al.\ 2001, \apjl, 556, 2, L91

\bibitem[Soker(2016)]{Soker2016} Soker, N.\ 2016, \nar, 75, 1

\bibitem[Stute \& Sahai(2009)]{Stute2009} Stute, M. \& Sahai, R.\ 2009, \aap, 498, 1, 209

\bibitem[Tejeda \& Toal{\'a}(2025)]{TejedaToala2025} Tejeda, E. \& Toal{\'a}, J.~A.\ 2025, \apj, 980, 2, 226

\bibitem[Toal{\'a}, Oskinova \& Vasquez-Torres(2025)]{Toala2025} Toal{\'a}, J.~A., Oskinova, L.~M., \& Vasquez-Torres, D.~A.\ 2025, , arXiv:2507.05203

\bibitem[Toal{\'a}(2024)]{Toala2024} Toal{\'a}, J.~A.\ 2024, \mnras, 528, 1, 987

\bibitem[Toal{\'a} et al.(2024)]{Toala2024_TCrB} Toal{\'a}, J.~A., Gonz{\'a}lez-Mart{\'\i}n, O., Sacchi, A., et al.\ 2024, \mnras, 532, 2, 1421

\bibitem[Toal{\'a} et al.(2023)]{Toala2023b} Toal{\'a}, J.~A., Gonz{\'a}lez-Mart{\'\i}n, O., Karovska, M., et al.\ 2023, \mnras, 522, 4, 6102

\bibitem[Toal{\'a}, Botello \& Sabin(2023)]{Toala2023a} Toal{\'a}, J.~A., Botello, M.~K., \& Sabin, L.\ 2023, \apj, 948, 1, 14

\bibitem[Toal{\'a} et al.(2022)]{Toala2022} Toal{\'a}, J.~A., Sabin, L., Guerrero, M.~A., et al.\ 2022, \apjl, 927, 1, L20

\bibitem[Torbett \& Campbell(1987)]{Torbett1987} Torbett, M.~V. \& Campbell, B.\ 1987, \apjl, 318, L29

\bibitem[Vander Meulen et al.(2023)]{VanderMeulen2023} Vander Meulen, B., Camps, P., Stalevski, M., et al.\ 2023, \aap, 674, A123

\bibitem[Vasquez-Torres et al.(2024)]{VasquezTorres2024} Vasquez-Torres, D.~A., Toal{\'a}, J.~A., Sacchi, A., et al.\ 2024, \mnras, 535, 3, 2724

\bibitem[Wheatley \& Kallman(2006)]{Wheatley2006} Wheatley, P.~J. \& Kallman, T.~R.\ 2006, \mnras, 372, 4, 1602

\bibitem[Whitelock(1987)]{Whitelock1987} Whitelock, P.~A.\ 1987, \pasp, 99, 573

\bibitem[Zamanov et al.(2024)]{Zamanov2024} Zamanov, R.~K., Stoyanov, K.~A., Marchev, V., et al.\ 2024, Astronomische Nachrichten, 345, e20240036

\bibitem[Zhekov \& Tomov(2019)]{Zhekov2019} Zhekov, S.~A. \& Tomov, T.~V.\ 2019, \mnras, 489, 2, 2930

\end{thebibliography}
\end{document}